# Th substituted SmFeAsO: structural details and superconductivity with $T_c$ above 50 K

N. D. Zhigadlo,[1] S. Katrych,[1] S. Weyeneth,[2] R. Puzniak,[2,3] P. Moll,[1] Z. Bukowski,[1] J. Karpinski,[1] H. Keller,[2] and B. Batlogg[1]

[1]*Laboratory for Solid State Physics, ETH Zurich, CH-8093 Zurich, Switzerland*

[2]*Physik-Institut der Universität Zürich, CH-8057 Zürich, Switzerland*

[3]*Institute of Physics, Polish Academy of Sciences, PL-02-668 Warsaw, Poland*

## Abstract

Superconducting poly- and single-crystalline samples of $Sm_{1-x}Th_xFeAsO$ with partial substitution of $Sm^{3+}$ by $Th^{4+}$ were synthesized and grown under high pressure and their structural, magnetic and transport properties are studied. The superconducting $T_c$ reaches values higher than 50 K. Bulk superconducting samples ($x$ = 0.08, 0.15, 0.3) do not show any signs of a phase transition from tetragonal to orthorhombic crystal structure at low temperatures. With Th substitution the unit cell parameters $a$ and $c$ shrink and the fractional atomic coordinate of the As site $z_{As}$ remains almost unchanged, while that of Sm/Th $z_{Sm/Th}$ increases. Upon warming from 5 K to 295 K the expansion of the FeAs layer thickness is dominant, while the changes in the other structural building blocks are smaller by a factor of ~ 1/5, and they compensate each other, since the As-Sm/Th distance appears to contract by about the same amount as the O-Sm/Th expands. The poly- and single-crystalline samples are characterized by a full diamagnetic response in low magnetic field, by a high intergrain critical-current density for polycrystalline samples, and by a critical current density of the order of $8 \times 10^5$ A/cm$^2$ for single crystals at 2 K in fields up to 7 T. The magnetic penetration depth anisotropy $\gamma_\lambda$ increases with decreasing temperature, a similar behavior to that of $SmFeAsO_{1-x}F_y$ single crystals. The upper critical field estimated from resistance measurements is anisotropic with slopes of ~5.4 T/K ($H||ab$ plane) and ~2.7 T/K ($H||c$ axis), at temperatures sufficiently far below $T_c$. The upper critical field anisotropy $\gamma_H$ is in the range of ~ 2, consistent with the tendency of a decreasing $\gamma_H$ with decreasing temperature, already reported for $SmFeAsO_{1-x}F_y$ single crystals.

E-mail:zhigadlo@phys.ethz.ch

## I. INTRODUCTION

Renewed interest in high-$T_c$ superconductivity research has been stimulated by the discovery of superconductivity in $LaFeAsO_{1-x}F_x$ ($T_c$ ~ 26 K) [1], followed by the quest of finding materials with higher $T_c$'s up to ~ 55 K [2-4] through replacing La with other rare earth elements (such as Ce, Pr, Nd, Sm, Gd, Tb, Dy, Ho, and Y) [2-10]. In all members of the so called "1111" family LnFePnO (Ln: lanthanide, Pn: pnictogen) the $Fe_2Pn_2$ layer is central for superconductivity, and the $Ln_2O_2$ layer plays a role as a charge-carrier source. A distinctive feature of this family is the possibility to vary the charge carrier's density and thus tuning of electronic properties through chemical substitution at different atomic sites. Electron doping can be realized by the partial substitutions of F for O [6] as well as by oxygen deficiency [11]. In addition to the carrier doping in $Ln_2O_2$ layers, the partial substitution of the Fe site with Co or Ni also leads to superconductivity [12, 13]. Although the valence of the doped Co and Ni ions seems to remain 2+, electron carriers are thought to be induced owing to the itinerant character of the 3*d* electrons [12]. Recent calculations of Wadati *et al.* [14] show that the extra *d* electrons are almost totally located within the muffin-tin sphere of the substituted site. It is suggested that Co and Ni act more like random scatters, scrambling the momentum space and washing out parts of the Fermi surface. Apart from chemical doping, superconductivity develops when "chemical pressure" is applied by partial substitution of smaller and isovalent P ions for As in LaFeAsO [15]. Finally, Ln in the 1111 type structure can also be easily substituted. "Hole-doped" superconductors were reported in the case of Sr-substituted $Ln_{1-x}Sr_xFeAsO$ (Ln = La, Pr, Nd) [16], while "electron-doped" superconductors were found through $Th^{4+}$ substitution at the lanthanide site in GdFeAsO [17], NdFeAsO [18], and LaFeAsO [19] with respective $T_c$'s up to 56 K, 38 K, and 30 K. The evolution of superconductivity in Th and F co-doped polycrystalline samples $Sm_{0.9}Th_{0.1}FeAsO_{1-y}F_y$ was reported too [20]. Besides playing a role of carrier's source, Th substitution is also interesting for the stabilization of 1111 phases with smaller lanthanides such as Tb, for example [21]. In this case the incorporation of relatively large $Th^{4+}$ ions leads to a correction of the lattice mismatch between $Ln_2O_2$ fluorite-type block layers and $Fe_2As_2$ layers. One can expect that thorium ions can serve as effective pinning centers, but it is not established yet. Obviously, it is important to carry out more investigations in this direction in order to extend the possibility to tune the electronic properties of the 1111 phase since it still holds the highest $T_c$'s among pnictides and looks promising for future practical applications [22]. Since superconducting properties are highly sensitive to local structure environments, investigations on single crystals are particularly suited to help in elucidating the anisotropic superconductivity in the LnFePnO system.

In this paper, we report the high-pressure synthesis and single crystal growth of superconducting $Sm_{1-x}Th_xFeAsO$ in which the carrier's density is increased by substituting $Th^{4+}$ for $Sm^{3+}$. The superconducting $T_c$ reaches values higher than 50 K. Detailed studies were carried out on polycrystalline samples with nominal Th content of $x = 0.3$ and $T_c$ of 51.5 K, and on single crystals with $T_c = 49.5$ K and $x = 0.11$, characterized by a full diamagnetic response in low magnetic fields and by a high critical current density at low temperatures in high magnetic fields exceeding 7 T.

## II. EXPERIMENTAL DETAILS

Polycrystalline samples and single crystals of $Sm_{1-x}Th_xFeAsO$ were produced in a cubic anvil high-pressure cell. Polycrystalline samples with nominal compositions of $Sm_{1-x}Th_xFeAsO$ ($x$ = 0, 0.08, 0.1, 0.15, 0.20, and 0.30) were prepared using SmAs, ThAs, $Fe_2O_3$, and Fe powders as starting materials. Single crystals of Th substituted SmFeAsO have been grown using $Sm_{0.7}Th_{0.3}FeAsO$ and $Sm_{0.65}Th_{0.35}FeAsO$ nominal compositions and NaCl/KCl as a flux. Though conditions of crystal growth still require further optimization, we were able to grow plate-like single crystals with a size of ~ 100 × 100 μm$^2$. The precursor to flux ratio was fixed to 1:1. The precursor powders were mixed and ground, and the pellets were pressed in a glove box filled with dry argon gas. The growth conditions were similar to those used for the growth of $SmFeAsO_{1-x}F_y$ single crystals [23]. Pellets containing precursor and flux were placed in a BN crucible inside a pyrophyllite cube with a graphite heater. The six tungsten carbide anvils generated pressure on the whole assembly. In a typical crystal growth run, a pressure of 3 GPa was applied at room temperature. While keeping the pressure constant, the temperature was ramped up to 1430 °C within 1 h, maintained for 65 h, and decreased in 1 h to room temperature. For the synthesis of polycrystalline samples the maximum temperature was maintained for 4.5 h, followed by quenching. Then the pressure was released, the sample extracted and in the case of crystal growth NaCl/KCl flux was dissolved in water.

Powder x-ray diffraction investigations (XRD) were performed at room temperature on a STOE diffractometer ($CuK_\alpha$ radiation, $\lambda$ = 1.54056 Å) equipped with a mini-PSD detector and a Ge monochromator on the primary beam. Three powder samples with different starting composition ($Sm_{1-x}Th_xFeAsO$, $x$ = 0.08, 0.15, and 0.3) were investigated at temperatures from

~ 5 K till 295 K, using synchrotron radiation (Swiss-Norwegian beam lines at the European Synchrotron Radiation Facility, Grenoble, France) and a mar345 image-plate area detector (the sample to detector distance of 250 mm was calibrated using a Si standard, $\lambda = 0.7000$ Å). Helium flow cryostats with a special chamber were used to reach low temperatures. Between individual runs the temperature was changed at a rate of ~ 5 K/min, and then the samples were allowed to equilibrate for 15 min before data collecting was started. The sample temperature was controlled with an accuracy of 1 K. The 2D diffraction images were integrated using the program fit2d [24] and the 1D powder pattern were refined with the FULLPROF program in a sequential mode [25]. Single crystal structural investigations were done at room temperature using an x-ray single-crystal Xcalibur PX, *Oxford* Diffraction diffractometer equipped with a charge coupled device (CCD) area detector (at the Laboratory of Crystallography, ETH Zurich), which allowed us to examine the whole reciprocal space (Ewald sphere). Data reduction and the analytical absorption correction were introduced using the CRYSALIS software package [26]. The crystal structure was determined by a direct method and refined on $F^2$, employing the SHELXS-97 and SHELXL-97 programs [27, 28].

Magnetic measurements were performed in a *Quantum Design* Magnetic Property Measurement System (MPMS XL) with the Reciprocating Sample Option (RSO) installed. The magnetic torque was measured using a highly sensitive miniaturized piezoresistive torque sensor within a home-made experimental setup described elsewhere [29, 30]. This technique allows to measure the angular dependent superconducting magnetic behavior by detecting the torque of a single crystal in a magnetic field along a chosen orientation with respect to the crystallographic *c*-axis. Four-point resistivity measurements were performed in a 14 Tesla *Quantum Design* Physical Property Measurement System (PPMS). To minimize the influence of material inhomogeneities, plate-like $Sm_{1-x}Th_xFeAsO$ crystals smaller that 100 × 100 μm$^2$ were selected and contacted using a Focused Ion Beam (FIB) method. With this technique micrometer-sized Pt leads are precisely deposited onto the crystal and was found not to alter the bulk superconducting properties [22].

## III. RESULTS AND DISCUSSION

### A. Crystal structure

Figure 1 depicts XRD patterns of the polycrystalline $Sm_{1-x}Th_xFeAsO$ samples with nominal composition of $x = 0$, 0.08, and 0.15. The parent compound ($x = 0$) is essentially single phase, and the main peaks of Th substituted samples ($x$ = 0.08 and 0.15) can be indexed based on a tetragonal cell ($P4/nmm$) of the ZrCuSiAs-type structure. According to the results of the Rietveld refinement [28] the impurity $ThO_2$ and SmAs phases in $Sm_{0.92}Th_{0.08}FeAsO$ and $Sm_{0.85}Th_{0.15}FeAsO$ samples amount to less than ~ 10 at.% (Fig. 1). For the samples substituted with more Th the amount of $ThO_2$ gradually increases. It is important to notice that in all of the reported syntheses [17-21] $ThO_2$ oxide was used as a Th source, and the final Th substituted GdFeAsO, NdFeAsO, LaFeAsO, $SmFeAsO_{1-y}F_y$, and TbFeAsO samples always contain $ThO_2$ as an impurity phase. To exclude or minimize the appearance of this parasitic phase we used ThAs instead of $ThO_2$ in the starting mixture, but we were not able to suppress the formation of the $ThO_2$ phase. This is a very stable oxide with a melting point of ~ 3660 K and once it is formed, it probably cannot be removed completely. Thus, the notations used for the nominal/initial compositions of the polycrystalline samples do not agree with the real composition, as we also quantify later in the structure analysis. Because the ionic radius of $Th^{4+}$ ($r_{\mathrm{Th}}^{\mathrm{ion}}$ ~ 0.95 Å) is smaller than that of $Sm^{3+}$ ($r_{\mathrm{Sm}}^{\mathrm{ion}}$ ~ 1.04 Å) [31] one can expect the lattice to shrink as a result of the Th substitution. Indeed, compared to SmFeAsO ($a$ = 3.9427(1) Å and $c$ = 8.4923(3) Å), both lattice parameters changed to $a$ = 3.9357(2) Å and $c$ = 8.4327(5) Å for $x$ = 0.3. Though the $a$-axis changes only slightly, the $c$-axis is shortening significantly with Th substitution, indicating that Th is indeed incorporated into the lattice. A similar trend in the shrinkage of the $c$-axis was already observed for the $Gd_{1-x}Th_xFeAsO$, $Nd_{1-x}Th_xFeAsO$, $La_{1-x}Th_xFeAsO$, and $Sm_{0.9}Th_{0.1}FeAsO_{1-y}F_y$ systems [17-20] and was attributed to the strengthening of the interlayer Coulomb attraction as a consequence of Th doping. On contrary, for the $Tb_{1-x}Th_xFeAsO$ system [21] both lattice parameters increase with increasing $x$ since the ionic size of $Th^{4+}$ is larger than that of $Tb^{3+}$ and in this case $Th^{4+}$ ions besides playing a role of carrier's source can also reduce the lattice mismatch between the $Ln_2O_2$ fluorite-type block layers and the $Fe_2As_2$ conducting layers.

In order to gain insight into the evolution of the crystal structure with Th substitution we performed low temperature powder XRD measurements for the samples with $x$ = 0.08, 0.15, and 0.3. It is well known that the parent SmFeAsO compound crystallizes in a $P4/nmm$ tetragonal structure at high temperatures, but undergoes a transformation to ($Cmma$) orthorhombic symmetry at low temperatures [32]. This transformation is accompanied by a spin density wave (SDW) formation, typical for the undoped LnFeAsO systems. However, upon substitution the SDW and structural transition are suppressed and superconductivity occurs. Several investigations have addressed this issue but there is controversy if the orthorhombic distortion survives when the material becomes superconducting (see, Ref. 33 and references therein). We addressed this question in the $Sm_{1-x}Th_xFeAsO$ system. Figure 2 shows low-temperature XRD patterns for three Th substituted samples ($x$ = 0.08, 0.15, and 0.3). By monitoring the gradual evolution of XRD patterns at low temperatures we paid special attention to the (220) peaks. No splitting or broadening of the (220) reflection was found down to liquid helium temperature, indicating no structural transformation. Even for lightly Th doped ($x$ = 0.08) bulk superconducting sample, we were not able to resolve any definite inflection point in the evolution of full width at half maximum (FWHM) with temperature, which could be a signature of a structural transformation. Therefore, our data support a picture of a rather abrupt suppression of the orthorhombic phase at the boundary to superconductivity, as observed, e.g., in $PrFeAsO_{1-x}F_x$ [34]. Single crystals right at the critical composition would remain interesting to study.

The refined structural data for $Sm_{1-x}Th_xFeAsO$ polycrystalline samples are summarized in Table 1. Since small amounts of $ThO_2$ and SmAs were observed as impurity phases they were included in the refinement models (using a multiphase Rietveld code) in order to minimize systematic errors and to obtain reliable structural parameters. The Rietveld refinement of the structure using powder x-ray diffraction data collected from liquid helium temperature to room temperature converged at $R_B$ = 1.1-2.3, $R_F$ = 1.2-2.1% for $x$ = 0.08; $R_B$ = 1.8-2.9, $R_F$ = 1.7-2.4% for $x$ = 0.15; $R_B$ = 2.0-2.3, $R_F$ = 1.6-1.9% for $x$ = 0.3; $R_p$ ~ 2.6%, $R_{wp}$ ~ 3.8% (not corrected for background); and $R_p$ ~ 7.0%, $R_{wp}$ ~ 8.3% (conventional Rietveld $R$-factors).

In Fig. 3 the temperature dependence is shown for the lattice parameters $a$ and $c$, the unit cell volume $V$, and of the relative thermal expansion $\Delta a/a$ and $\Delta c/c$, where error bars are not always visible since the uncertainty in the determination of quantities is smaller than the size of the symbols. The coefficients of thermal expansion around room temperature were estimated graphically. The linear thermal expansion coefficients $\alpha_a$ and $\alpha_c$ vary not much with variation of Th content, however the thermal expansion is strongly anisotropic. At 295 K, $\alpha_c$ is about four times larger than $\alpha_a$ ($21.7 \times 10^{-6}$ and $6.0 \times 10^{-6}$ $K^{-1}$, respectively). The thermal expansion along the $c$ axis can be seen as a sum of the changes in the FeAs layer, the (Sm,Th)O layer, and their separation [(Sm,Th)As] distance. Remarkably, it is an increase of the Fe-As layer thickness that dominates the expansion, while the changes in the other structural building blocks are smaller by a factor of ~ 1/5, and they compensate each other, since the As-Sm/Th appears to contract as much as the O-Sm/Th expands upon warming from 5 K to 295 K. The anisotropic behavior of the thermal expansion is less pronounced at lower temperatures (Fig. 3). Within the present resolution, the cell parameters vary continuously across the superconducting transition.

To document the effect of Th substitution in more detail we carried out single crystal x-ray diffraction studies. Individual plate-like crystals (~ 100 × 100 $\mu m^2$) were selected. The quality of the crystals was checked by means of XRD using an x-ray diffractometer equipped with a CCD area detector, which allows us to examine the entire reciprocal space (Ewald sphere) for the presence of other phases or crystallites with different orientation. Crystals from various batches show well-resolved reflection patterns, indicating a high quality of the crystal structure. As clearly seen in Fig. 4, no additional phases (impurities, twins or intergrowing crystals) were detected by examining the reconstructed reciprocal space sections. The crystallographic and structural refinement parameters of SmFeAsO and $Sm_{1-x}Th_xFeAsO$ single crystals are summarized in Table 2. Since previous results [11] for the "1111" family suggested that oxygen non-stoichiometry (deficiency) can induce superconductivity, we allowed in the refinement the occupancies of all sites to vary. Within experimental error, all occupancies are equal to one. From the refinement analysis, the Th content for the batch with starting composition $Sm_{0.65}Th_{0.35}FeAsO$ is only about 11 at.%. As for polycrystalline samples, the values of both the *a*-axis and *c*–axis lattice parameters of the parent compound are significantly larger than those of Th substituted. For 11 at.% substitution the relative changes, $\Delta c/c$ ~ -0.49% and $\Delta a/a$ ~ -0.15%. Lee *et al.* [35] pointed out that there is a relationship between $T_c$ and the As-Fe-As bond angle of the $FeAs_4$-tetrahedron; $T_c$ is maximal when the As-Fe-As bond angle is close to 109.47, corresponding to an ideal tetrahedron. Several groups [36] studied it theoretically, and later Mizuguchi *et al.* [37] investigated it experimentally and finally it was suggested that the pnictogen height ($h_{Pn}$) is an important parameter for controlling $T_c$ and the nesting properties of Fermi surfaces. This geometry parameter $h_{Pn}$ measures the distance between a pnictogen atom and the Fe layer, given numerically by ($z_{Pn}$ – 0.5) × *c*, where $z_{Pn}$ is the internal coordinate of the pnictogen atom and *c* the *c*-axis lattice constant. As it was shown in [36] $z_{Pn}$ controls the relative position of the $d_{x2-y2}(=d_{xy})$ and $d_{z2}$ bands near the Γ point of the Brillion zone. All together these findings strongly indicate that the superconducting properties of the Fe-based superconductors are closely correlated with the crystal structure. Our own detailed structure information obtained from single crystal refinement shows that the As-Fe-As bond angles, α and β, are very close to those of a regular tetrahedron and they are almost unchanged by Th substitution. The value of $h_{Pn}$ ~ 1.37 Å is very close to its optimal value and almost insensitive to Th substitution (see, Table 1 and 2). Similar results were found for F doped SmFeAsO single crystals. Bearing all this in mind we conclude that intrinsically the geometry of $FeAs_4$ tetrahedron in SmFeAsO is close to optimal for having a high $T_c$, and the geometry does not change significantly when we introduce charge carriers through heterovalent substitution in the SmO layer. While the substitutions add electrons, the $FeAs_4$ tetrahedron keeps the optimal geometry.

The effect of Th substitution details is shown schematically in Fig. 5a. With increasing Th content the main metric parameters vary systematically: both unit cell parameters ($a$, $c$) shrink, the fractional atomic coordinate of the As site $z_{As}$ remains almost unchanged, while that of Sm/Th $z_{Sm/Th}$ increases. The As-Sm/Th distance shortens and the O-Sm/Th distance expands (Tables 1 and 2). Equivalently one may focuses on the "layers" of the structure: the Sm/ThO layer expands ($\Delta y$ = 0.055 Å), the AsFe layer remains unaffected, and the distance $x$ between the Sm/ThO and the AsFe layers shortens by $\Delta x$ = -0.048 Å (Table 2). These changes can be rationalized by considering the degree of covalent and ionic bonding character [38], and the fact that the ionic radius of tetravalent Th is smaller than that of trivalent Sm whereas Th has a larger covalent radius than that of Sm ($r_{\mathrm{Th}}^{\mathrm{cov}}$ ~ 1.65 Å, $r_{\mathrm{Sm}}^{\mathrm{cov}}$ ~ 1.62 Å). The same geometric changes of the unit cell due to Th doping were observed for polycrystalline samples (Table 1). Th substitution influences the $FeAs_4$ tetrahedron size less than the variation with temperature (Fig. 5b) since the $FeAs_4$ tetrahedron expands considerably along the $c$ direction with increasing temperature. The angle $\alpha$ ($\beta$) is getting smaller (larger), and $h_{Pn}$ is getting larger for $Sm_{0.7}Th_{0.3}FeAsO$ in the temperature range from 15 to 295 K: $\Delta\alpha = -1$ °, $\Delta\beta = 0.4$ °, $\Delta h_{Pn}$ = 0.02 Å. The pnictogen height as a function of temperature for three Th composition is shown in Fig. 5c. It varies in the same manner for all studied samples from ~1.37 Å at 295 K to ~1.35 Å at 5 K.

### B. Magnetic measurements and torque magnetometry

Figure 6 shows the temperature dependence of the magnetic susceptibility of $Sm_{1-x}Th_xFeAsO$ polycrystalline samples in a magnetic field of 0.5 mT, measured after zero-field-cooling (ZFC). The sample with $x$ = 0.08 shows bulk superconductivity with $T_c$ ~ 38 K. The inhomogeneity of the Th distribution at low levels of doping leads to a broadening of the transition. With increasing Th content the samples become more homogeneous, and the superconducting transition shifts to higher temperatures, reaching a maximum onset transition temperature of ~ 51.5 K for $x$ = 0.3.

Initial magnetization curves recorded for polycrystalline $Sm_{0.7}Th_{0.3}FeAsO$ in the temperature range between 2 and 40 K indicate that the intergranular lower critical field $H_{c1}$ is very high (Fig. 7), indicative for good connections between the grains. This is confirmed by results of minor hysteresis loops measurements, performed in the zero-field-cooled mode, in the field range before flux is trapped within the grains [39]. In Fig. 8 the intergrain critical current density is shown for polycrystalline $Sm_{0.7}Th_{0.3}FeAsO$ estimated in the temperature range from 2 to 40 K applying Bean's model [40], with the sample size in the plane perpendicular to the applied field direction. As shown in Fig. 9 the irreversibility line for polycrystalline $Sm_{0.7}Th_{0.3}FeAsO$ is well described by a power-law temperature dependence according to $(1-T/T_c)^n$ with a $T_c = 51.4(1)$ K, and $n = 1.56(2)$ (straight line in the log-log $H_{irr}$ vs. $(1-T/T_c)$ dependence in the inset). The value of $n = 1.56(2)$ is very close to $n = 3/2$, typically obtained for high-$T_c$ superconductors characterized by a small coherence length [41].

Figure 10 shows the normalized diamagnetic magnetization measured on small lump samples which were mechanically extracted from the single crystal growth products. The highest onset $T_c = 53$ K was observed for $Sm_{0.65}Th_{0.35}FeAsO$ starting composition, being comparable to the maximum $T_c$ reported for F doping [2-4]. Since lump samples include, in addition to the desired "1111" phase, also flux and impurity phases, the magnetic susceptibility in the normal state shows a positive paramagnetic contribution. Temperature and compositional gradients in the crucible may lead to variations in the Th content of the crystals and differences in $T_c$ giving rise to a broad over-all onset of diamagnetism. The temperature dependence of the magnetic moment measured in a magnetic field parallel to the *c*-axis of a plate-like single crystal of rectangular shape with mass of 100 ng and a crystallographically determined composition of $Sm_{0.89}Th_{0.11}FeAsO$ is shown in Fig. 11. A sharp transition to the superconducting state with $T_{c,eff} = 49.5$ K is observed.

Single crystalline samples may be studied for their anisotropic superconducting properties by analyzing the magnetic torque $\tau(\theta) = \mu_0 mH \sin(\theta)$, where $\theta$ denotes the angle between the applied magnetic field and the *c*-axis of the crystal, $\mu_0$ is the magnetic constant, $m$ the angular dependent magnetic moment of the sample, and $H$ the magnitude of the applied magnetic field. Measurements of the magnetic torque, similar to those carried out on fluorine substituted SmFeAsO single crystals [30, 42] were performed in order to investigate the anisotropic superconducting behavior in terms of the anisotropy parameter $\gamma$. The torque data were analyzed using a theoretical expression derived by Kogan *et al.* [43] within the London approximation of anisotropic Ginzburg-Landau theory:

$$\tau(\theta) = -\frac{V\Phi_0 H}{16\pi\lambda_{ab}^2}\left(1-\frac{1}{\gamma^2}\right)\frac{\sin(2\theta)}{\varepsilon(\theta)}\ln\left(\frac{\eta H_{c2}^{c}}{\varepsilon(\theta)H}\right) + A\sin(2\theta). \tag{2}$$

Here V is the volume of the crystal, $\Phi_0$ is the elementary flux quantum, $H_{c2}^{\|c}$ is the upper critical field along the *c*-axis of the crystal, η denotes a numerical parameter of the order of unity, and $\varepsilon(\theta) = \left[\cos^2(\theta) + \gamma^{-2}\sin^2(\theta)\right]^{1/2}$. The volume of the crystal given in Fig. 11 was estimated from magnetization measurements in low magnetic fields applied along the samples′ *ab*-plane. The *c*-axis dimension is much smaller than the *a*- and *b*- dimensions, wherefore demagnetizing effects in this geometry can be neglected. The crystal dimensions agree well with measurements using a optical microscope. It was discussed in earlier investigations on F-substituted SmFeAsO that the anisotropy parameter γ, determined from torque measurements in fields much below the upper critical field, mainly probes the magnetic penetration depth anisotropy $\gamma_\lambda$ [30]. The experimental setup as well as the measurement procedure is described in detail elsewhere [29, 30]. Representative examples of angular dependent torque curves measured at 45 K in a magnetic field of 1.2 T are shown in Fig. 12. The presence of a finite torque as well as a strongly angular dependent irreversibility clearly reveals the pronounced anisotropy of the superconducting state. A pronounced irreversibility close to the *ab*-plane (90 degrees) is observed, associated with a very high vortex-pinning in this field direction with respect to the crystallographic orientation. The peak-like feature very close to 90 degrees is interpreted as an indication of the lock-in effect of vortices within the *ab*-plane, indicative of a high crystal quality. If imperfection or impurity pinning effects dominate, the lock-in effect cannot be resolved, usually ascribed to intrinsic pinning of vortices due to the anisotropic crystal structure.

Note that the anisotropy parameter γ derived from the present analysis in rather low magnetic fields equals the magnetic penetration depth anisotropy $\gamma_\lambda$ [30]. The derived results of $\gamma_\lambda$ are shown in Fig. 13, which increase from ~ 9.8 at 47 K to ~ 13.2 at 38 K. Unfortunately, due to the pronounced vortex-pinning in this crystal, especially close to the *ab*-plane, torque data below 38 K could not be reliably analyzed. The observed temperature dependence of $\gamma_\lambda$ is in good agreement with $\gamma_\lambda$(T) found for fluorine substituted SmFeAsO and NdFeAsO [30, 42]. Obviously, the $Sm_{1-x}Th_xFeAsO$, $SmFeAsO_{1-x}F_y$, and $NdFeAsO_{1-x}F_y$ are much more anisotropic superconductors than e.g. $BaFe_2As_2$, $SrFe_2As_2$, and $FeSe_{0.5}Te_{0.5}$, with $\gamma_\lambda$ ~ 2-3 [44-49].

Magnetization hysteresis loops of a relatively large single crystal of $Sm_{1-x}Th_xFeAsO$ (~ 500 ng) measured at various temperatures below $T_c$ in magnetic fields up to 7 T applied parallel to the crystal *c*-axis are presented in Fig. 14 (upper panel). The wide loops, with a width almost independent of the field, indicate a rather high critical current density ($J_c$) in the sample. The $J_c$ estimated from the width of the hysteresis loop using Bean's model is at 2 K close to $10^6$ A $cm^{-2}$ in the field range investigated (see Fig. 14, lower panel). The slight increase in critical current density for higher magnetic fields may indicate an increase in the effectiveness of pinning centers giving rise to a "peak effect". A similar behavior was found in F doped SmFeAsO single crystals [22, 23, 38].

### C. Magnetoresistance of $Sm_{1-x}Th_xFeAsO$

A Focus Ion Beam (FIB) based contacting method [22] was employed to contact $Sm_{1-x}Th_xFeAsO$ single crystals. The insert to the upper panel of Fig. 15 shows a typical cool-down curve. The resistivity in zero field decreases by a factor of 3.7 from room temperature upon cooling to $T_c$ (onset - 49.7 K; 50% - 48.9 K; zero – 47.4 K), defining the so-called resistivity ratio of the sample studied here. Similar resistivity ratios are typically observed in SmFeAsO with various substitutions on the Fe, As or O sites [22, 38, 50]. A very slight change of slope near 125 K is observed, which is possibly related to the fluctuations of the spin density wave instabilities. The magnetoresistance in fixed fields was recorded upon decreasing temperature starting from above $T_c$, to account for irreversibility effects (Fig. 15). From this data, the upper critical field was estimated, defined as the magnetic field, where 50% of the resistivity is suppressed of $\rho_n$ (see inset in lower panel of Fig. 15), where $\rho_n$ is the linear extrapolation of the normal state resistivity. In $Sm_{1-x}Th_xFeAsO$ magnetic fields cause only a slight shift of the onset of superconductivity, but a significant broadening of the transition, indicating weaker pinning and accordingly larger flux flow dissipation. Such behavior has been widely observed in the 1111 family. Interestingly, a similar broadening of the resistivity with $H||c$ was also observed in cuprate superconductors [51] and was interpreted in terms of a vortex-liquid state [52, 53]. A recent report on $NdFeAsO_{1-x}F_x$ single crystals confirmed the existence of a vortex-liquid state in the 1111 system [54]. Using a linear part of $H_{c2}(T)$, the upper critical field slopes $dH_{c2}/dT$ ~ 5.4 T/K for $H||ab$ and ~ 2.7 T/K for $H||c$ were determined. These slopes suggest very high values of $H_{c2}(0)$. The ratio $dH_{c2}/dT_{(H||ab)}/dH_{c2}/dT_{(H||c)}$ provides a rough estimation of the upper critical field anisotropy $\gamma_H$ for the temperatures significantly below $T_c$ leading, for the slopes values determined here, to a value of $\gamma_H(T = 0) \approx 2$. The results suggest a tendency of decreasing $\gamma_H$ with decreasing temperature, as already reported from high field resistivity measurements for $NdFeAsO_{1-x}F_x$ [55] and $SmFeAsO_{1-x}F_y$ [38]. Such a small value of the ratio $dH_{c2}/dT_{(H||ab)}/dH_{c2}/dT_{(H||c)}$ is mostly due to the enormous sensitivity of the steep $dH_{c2}/dT_{(H||ab)}$ on Th doping. Although very similar in shape, its value is ~ 5.4 T/K in $Sm_{1-x}Th_xFeAsO$ and ~ 8 T/K in $SmFeAsO_{1-x}F_y$ [38]. Furthermore, $H_{c2}$ for $H||c$ is steeper in $Sm_{1-x}Th_xFeAsO$.

## IV. CONCLUSIONS

Poly- and single-crystalline samples of $Sm_{1-x}Th_xFeAsO$ superconductors were successfully prepared using the high-pressure cubic anvil technique, and their crystallographic and basic superconducting state properties were studied. Partial substitution of $Sm^{3+}$ by $Th^{4+}$ results in a decrease of the unit cell volume and the appearance of bulk superconductivity with onset $T_c$ higher than 50 K. Upon warming from 5 K to 295 K the over-all unit cell expands significantly more perpendicular to the layers than parallel to them. Remarkably, it is the increase of the FeAs layer thickness that dominates the expansion, while the changes in the other layers are smaller and even compensate each other, since the As-Sm/Th distance appears to contract by about the same amount as the O-Sm/Th distance expands. Magnetic measurements performed on a plate-like crystal with a sizes of ~ 100 × 100 μm$^2$ and $T_c$ = 49.5 K show a relatively high critical current density of $10^6$ A/cm$^2$ at 2 K almost independent of the magnetic field. The magnetic penetration depth anisotropy $\gamma_\lambda$ increases with decreasing temperature. The upper critical fields $H_{c2}$ in $Sm_{1-x}Th_xFeAsO$ extracted from the resistivity measurements is anisotropic with slopes of ~ 5.4 T/K ($H||ab$ plane) and ~ 2.7 T/K ($H||c$ axis), sufficiently far below $T_c$. The upper critical field anisotropy $\gamma_H$ (very roughly estimated for the temperatures far below $T_c$) is in the range of ~ 2, consistent with the tendency of $\gamma_H$ to decrease with decreasing temperature, as already reported for other $NdFeAsO_{1-x}F_x$ and $SmFeAsO_{1-x}F_y$ compounds. All together, the unusual temperature behavior of both anisotropy parameters, $\gamma_\lambda$ and $\gamma_H$, observed in Th substituted SmFeAsO further support a common multi-gap scenario proposed for FeAs-based superconductors [30, 38, 42, 56].

## ACKNOWLEDGMENTS

We thank Drs. D. Chernyshov and Y. Filinchuk for their assistance at the Swiss-Norwegian Beamlines, ESFR, Grenoble, France. We also thank Prof. W. Steuer (ETH) for the use of the x-ray single-crystal diffractometer. This work was supported by the Swiss National Science Foundation through the National Center of Competence in Research MaNEP (Materials with Novel Electronic Properties), project No.124616 and the Polish Ministry of Science and Higher Education, within the research project for the years 2007-2010 (No. N N202 4132 33).

**TABLE 1.** Refined structural parameters for $Sm_{1-x}Th_xFeAsO$ samples measured at 15 and 295 K. The lattice is tetragonal with space group *P*4/*nmm*.

| Starting composition | | $Sm_{0.7}Th_{0.3}FeAsO$ | $Sm_{0.85}Th_{0.15}FeAsO$ | $Sm_{0.92}Th_{0.08}FeAsO$ |
|---|---|---|---|---|
| Data collection temperature (K) | | | 15 | |
| $T_c$ (K) | | 51.5 | 42 | 38 |
| *a* (Å) | | 3.9357(2) | 3.9381(3) | 3.9391(2) |
| *c* (Å) | | 8.4327(5) | 8.4475(3) | 8.4540(4) |
| *V* (Å$^3$) | | 130.62(1) | 131.00(2) | 131.18(1) |
| Sm/Th *z* | | 0.1423(3) | 0.1393(2) | 0.1387(2) |
| $B_{iso}$ [1](Å$^2$) | | 1.31(9) | 3.85(6) | 1.68(9) |
| *Occupation* Sm/Th | | 0.88/0.12(4) | 0.92/0.08(3) | 100/0(2) |
| Fe $B_{iso}$ (Å$^2$) | | 1.3(2) | 3.9(1) | 2.2(2) |
| $z_{As}$ | | 0.6602(6) | 0.6603(4) | 0.6597(5) |
| $B_{iso}$ (Å$^2$) | | 1.8(2) | 4.4(1) | 2.2(2) |
| O $B_{iso}$ (Å$^2$) | | 5.3(9) | 1.8(4) | 1.8(6) |
| $R_p$ (%) | | 2.51[2](6.91[3]) | 2.06(7.49) | 1.66(7.31) |
| $R_{wp}$ (%) | | 3.77(8.27) | 2.68(6.91) | 2.31(7.12) |
| Sm/Th-As (4x) (Å) | | 3.246(3) | 3.258(2) | 3.266(2) |
| Sm-O (4x) (Å) | | 2.305(1) | 2.2937(8) | 2.2918(9) |
| Fe-As (4x) (Å) | | 2.384(3) | 2.391(2) | 2.388(2) |
| Fe-Fe (4x) (Å) | | 2.78299(2) | 2.78469(2) | 2.78539(1) |
| $As_1$-Fe-$As_2$, $\beta$ (deg) | | 108.6(3) | 108.8(2) | 108.6(2) |
| $As_2$-Fe-$As_3$, $\alpha$ (deg) | | 111.3(1) | 110.90(6) | 111.14(8) |
| *x* (Å) | | 1.666(6) | 1.692(4) | 1.704(4) |
| *y* (Å) | Fig. 5a | 2.399(5) | 2.354(3) | 2.345(4) |
| $z/h_{pn}$ (Å) | | 2.70(1)/1.351(6) | 2.708(6)/1.354(3) | 2.700(8)/1.350(4) |
| Data collection temperature, K | | | 295 | |
| *a* (Å) | | 3.9404(2) | 3.9433(3) | 3.9427(1) |
| *c* (Å) | | 8.4730(6) | 8.4810(6) | 8.4885(3) |
| *V* (Å$^3$) | | 131.56(1) | 131.87(2) | 131.955(8) |
| Sm/Th *z* | | 0.1421(3) | 0.1389(2) | 0.1385(2) |
| $B_{iso}$ [1](Å$^2$) | | 1.6(1) | 3.54(6) | 2.06(7) |
| *Occupation* Sm/Th | | 0.90/0.10(5) | 0.92/0.08(3) | 100/0(2) |
| Fe $B_{iso}$ (Å$^2$) | | 1.9(2) | 3.8(1) | 2.3(1) |
| $z_{As}$ | | 0.6618(7) | 0.6612(3) | 0.6609(4) |
| $B_{iso}$ (Å$^2$) | | 2.4(2) | 4.1(1) | 2.6(1) |
| O $B_{iso}$ (Å$^2$) | | 3.6(9) | 1.8(4) | 1.1(5) |
| $R_p$ (%) | | 2.34[2](7.57[3]) | 1.87(7.30) | 1.00(6.66) |
| $R_{wp}$ (%) | | 3.66(9.17) | 2.39(6.51) | 1.38(5.77) |
| Sm/Th-As (4x) (Å) | | 3.244(3) | 3.263(2) | 3.267(2) |
| Sm-O (4x) (Å) | | 2.309(1) | 2.2969(7) | 2.2955(9) |
| Fe-As (4x) (Å) | | 2.400(3) | 2.399(1) | 2.398(2) |
| Fe-Fe (4x) (Å) | | 2.78632(2) | 2.78830(2) | 2.787932(8) |
| $As_1$-Fe-$As_2$, $\beta$ (deg) | | 109.0(3) | 108.9(1) | 108.9(2) |
| $As_2$-Fe-$As_3$, $\alpha$ (deg) | | 110.3(1) | 110.52(5) | 110.60(7) |
| *x* (Å) | | 1.662(6) | 1.697(3) | 1.702(4) |
| *y* (Å) | Fig. 5a | 2.408(5) | 2.356(3) | 2.351(3) |
| $z/h_{pn}$ (Å) | | 2.74(1)/1.371(6) | 2.731(6)/1.3654(3) | 2.732(6)/1.366(3) |

[1] Debye-Waller factor; [2] not corrected for background
[3] conventional Rietveld *R*-factors

**TABLE 2**. Crystallographic and structure refinement parameters of SmFeAsO and $Sm_{1-x}Th_xFeAsO$ single crystals. The diffraction study is performed at 295(2) K using $MoK_\alpha$ radiation with $\lambda$ = 0.71073 Å. The lattice is tetragonal, *P*4/*nmm* space group with *Z* = 2, atomic coordinates: Sm/Th on 2*c* (1/4, 1/4, $z_{Sm}$), Fe on 2*b* (3/4, 1/4, 1/2), As on 2*c* (1/4, 1/4, $z_{As}$), O on 2*a* (3/4, 1/4, 0). The absorption correction was done analytically. A full-matrix least-squares method was employed to optimize $F^2$. Some distances and marking of atoms are shown in Fig. 5.

| Empirical formula | SmFeAsO (AN631_1) | $Sm_{0.89}Th_{0.11}FeAsO$(AN709_2) |
|---|---|---|
| $T_c/\Delta T_c$ (K) | – | 49.5/2 |
| Unit cell dimensions (Å) | *a* = 3.9427(1), *c* = 8.4923(3) | *a* = 3.9369(1), *c* = 8.4510(6) |
| Volume (Å$^3$) | 132.012(7) | 130.984(10) |
| $z_{Sm}$ (atomic coordinate) | 0.1372(1) | 0.1411(1) |
| *Occupation* Sm/Th | 1/0 | 0.89/0.11 |
| $z_{As}$ | 0.6603(1) | 0.6611(1) |
| $Sm_1$-$Sm_2$ (Å) | 3.6340(3) | 3.6658(7) |
| $Sm_1$-$Sm_3$(*a*) (Å) | 3.9427(1) | 3.9369(1) |
| O-O= Fe-Fe (Å) | 2.7879 | 2.7838 |
| $Sm_2$-$As_2$ (Å) | 3.2756(5) | 3.2469(9) |
| Sm-O | 2.2901(2) | 2.3015(4) |
| $As_2$-$As_3$(*a*) (Å) | 3.9427(1) | 3.9369(1) |
| $As_1$-$As_2$ (Å) | 3.8963(9) | 3.8945(15) |
| Fe-As (Å) | 2.3955(6) | 2.3937(9) |
| $As_1$-Fe-$As_2$, $\beta$ (deg) | 108.83(2) | 108.89(3) |
| $As_2$-Fe-$As_3$, $\alpha$ (deg) | 110.76(4) | 110.64(6) |
| *x* (Å) | 1.720(1) | 1.672(1) |
| *y* (Å) see, Fig. 5a | 2.330(2) | 2.385(2) |
| $z/h_{pn}$ (Å) | 2.723(2)/1.3613(9) | 2.723(2)/1.3615(9) |
| Calculated density (g/cm$^3$) | 7.475 | 7.761 |
| Absorption coefficient (mm$^{-1}$) | 39.606 | 43.713 |
| *F*(000) | 258 | 264 |
| Crystal size ($\mu$m$^3$) | 137 x 90 x 18 | 71 x 67 x 18 |
| $\Theta$ range for data collection (deg) | 4.80 to 47.11 | 4.82 to 45.88 |
| Index ranges | $-6<=h<=8$, $-8<=k<=7$, $-17<=l<=17$ | $-6<=h<=7$, $-6<=k<=7$, $-16<=l<=13$ |
| Reflections collected/unique | 3106/399 $R_{int.}$= 0.0643 | 2229/376 $R_{int.}$= 0.0457 |
| Completeness to 2$\Theta$ | 98.5% | 98.9% |
| Data/restraints/parameters | 399/0/11 | 376/0/12 |
| Goodness-of-fit on $F^2$ | 1.057 | 1.080 |
| Final *R* indices [$I$>2$\sigma$($I$)] | $R_1$ = 0.0383, w$R_2$ = 0.0921 | $R_1$ = 0.0454, w$R_2$ = 0.1030 |
| *R* indices (all data) | $R_1$ = 0.0402, w$R_2$ = 0.0933 | $R_1$ = 0.0591, w$R_2$ = 0.1049 |
| $\Delta\rho_{max}$ and $\Delta\rho_{min}$ (e/A$^3$) | 9.459 and -5.773 | 12.052 and -3.329 |

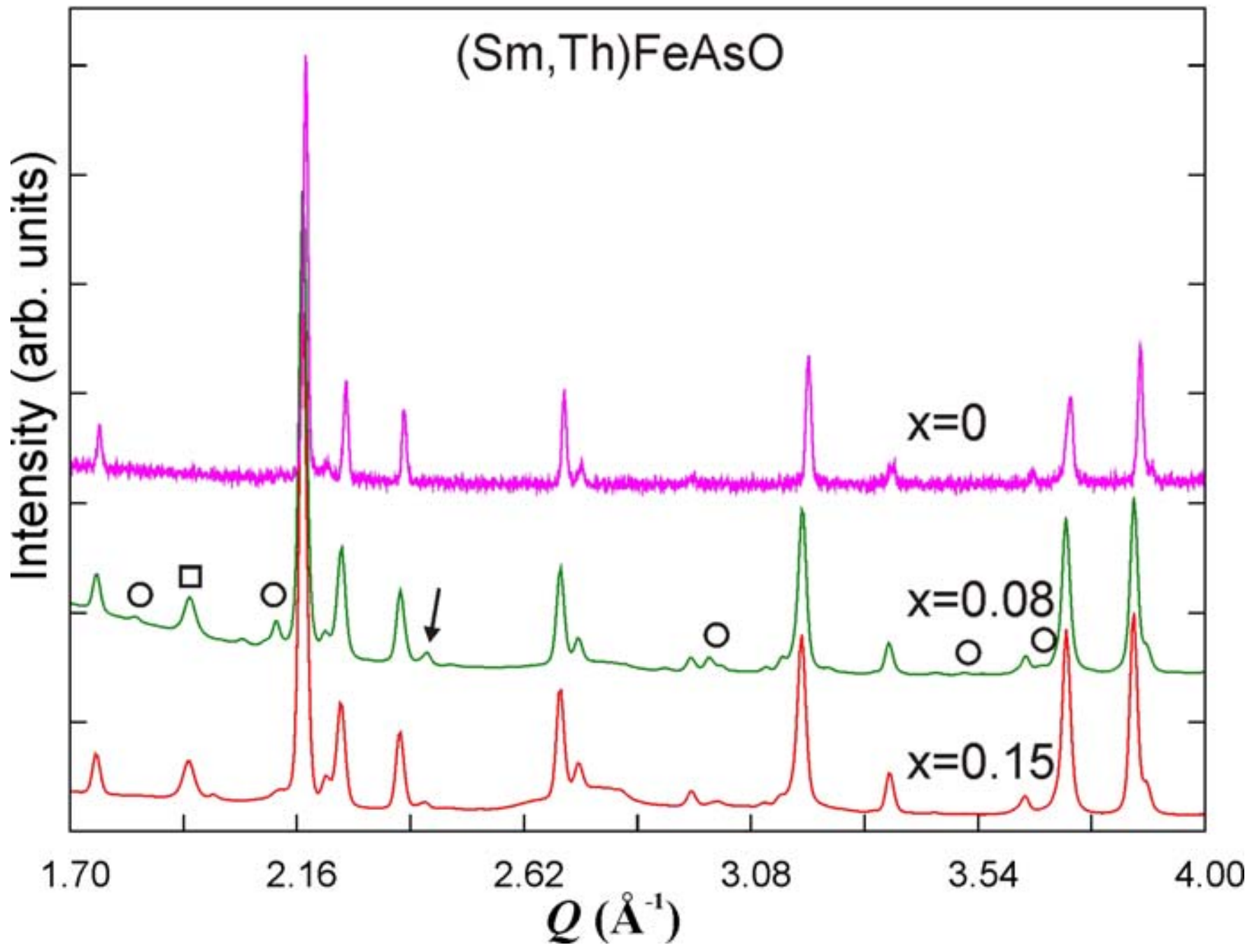

**FIG. 1.** (Color online) XRD patterns for $Sm_{1-x}Th_xFeAsO$ ($x$ = 0, 0.08, and 0.15) polycrystalline samples. The peaks marked by circles and square are due to the SmAs and $ThO_2$, respectively. The arrow marks scattering from the sample container.

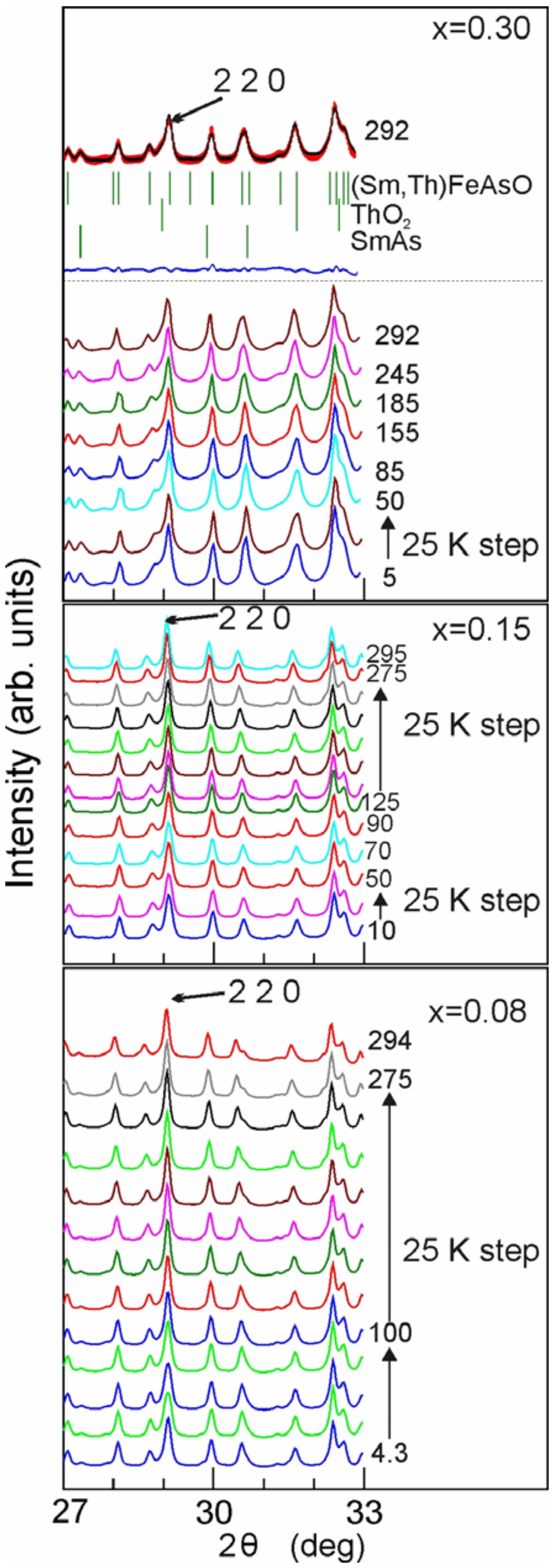

**FIG. 2.** (Color online) XRD patterns for $Sm_{1-x}Th_xFeAsO$ samples with $x$ = 0.30, 0.15, and 0.08. On the top of upper panel the red and black lines represent the experimental data and calculated profile, the blue line is the difference curve, green bars mark the reflections position of the phases constituting the samples. Temperatures in Kelvin are given on the right site.

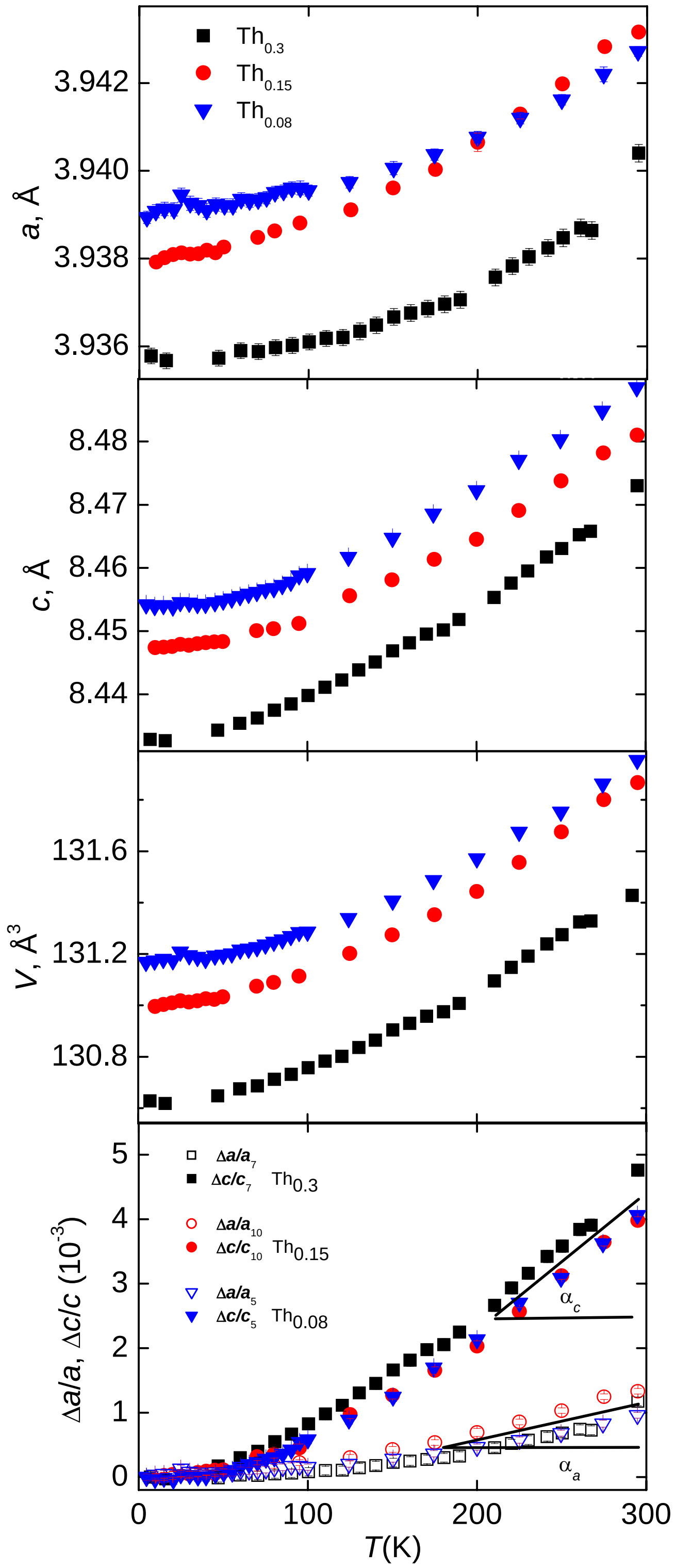

**FIG. 3.** (Color online) Temperature dependence of the structure parameters $a$ and $c$, elementary cell volume $V$, and of the relative thermal expansions $\frac{\Delta a}{a_T}, \frac{\Delta c}{c_T}$ for the three $Sm_{1-x}Th_xFeAsO$ compositions. The thermal expansion coefficient $\alpha_c$ is about four times larger than $\alpha_a$ ($21.7 \times 10^{-6}$ and $6.0 \times 10^{-6} K^{-1}$, respectively).

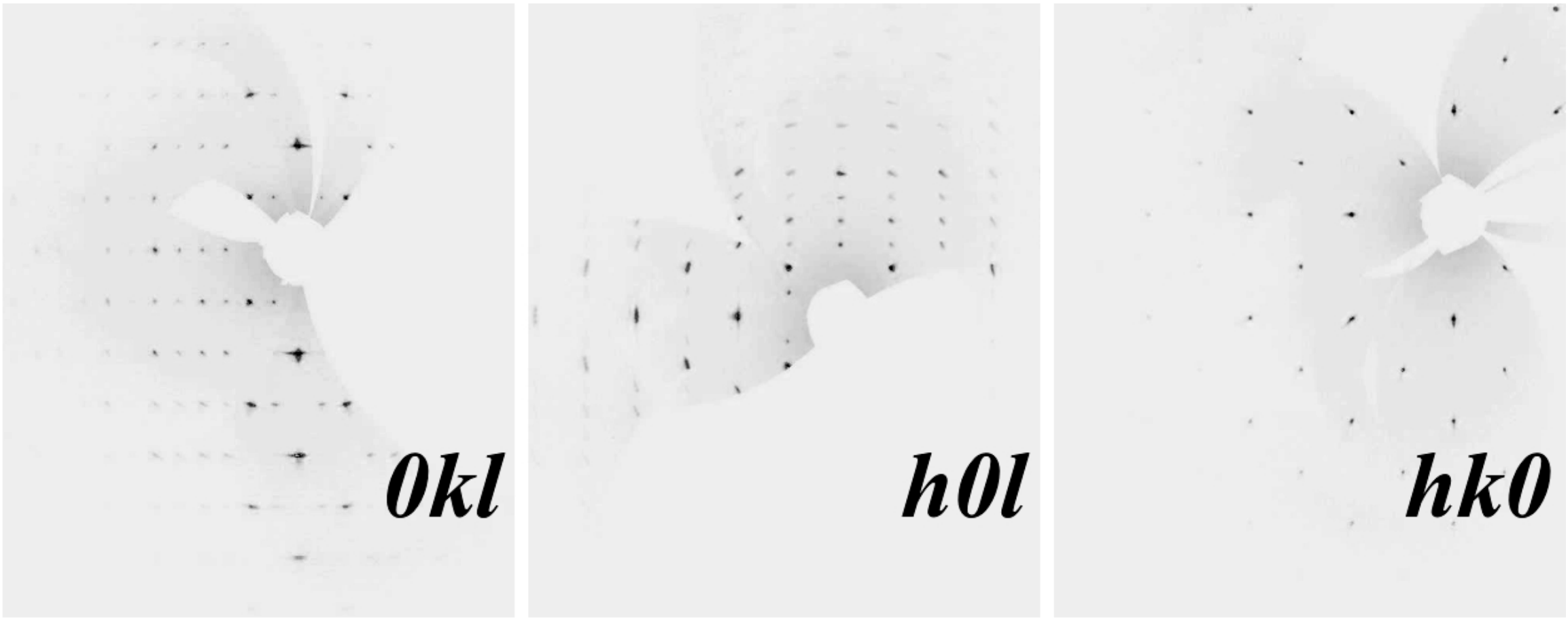

**FIG. 4.** The reconstructed *0hk*, *h0l*, and *hk0* reciprocal space sections of the $Sm_{0.89}Th_{0.11}FeAsO$ single crystal.

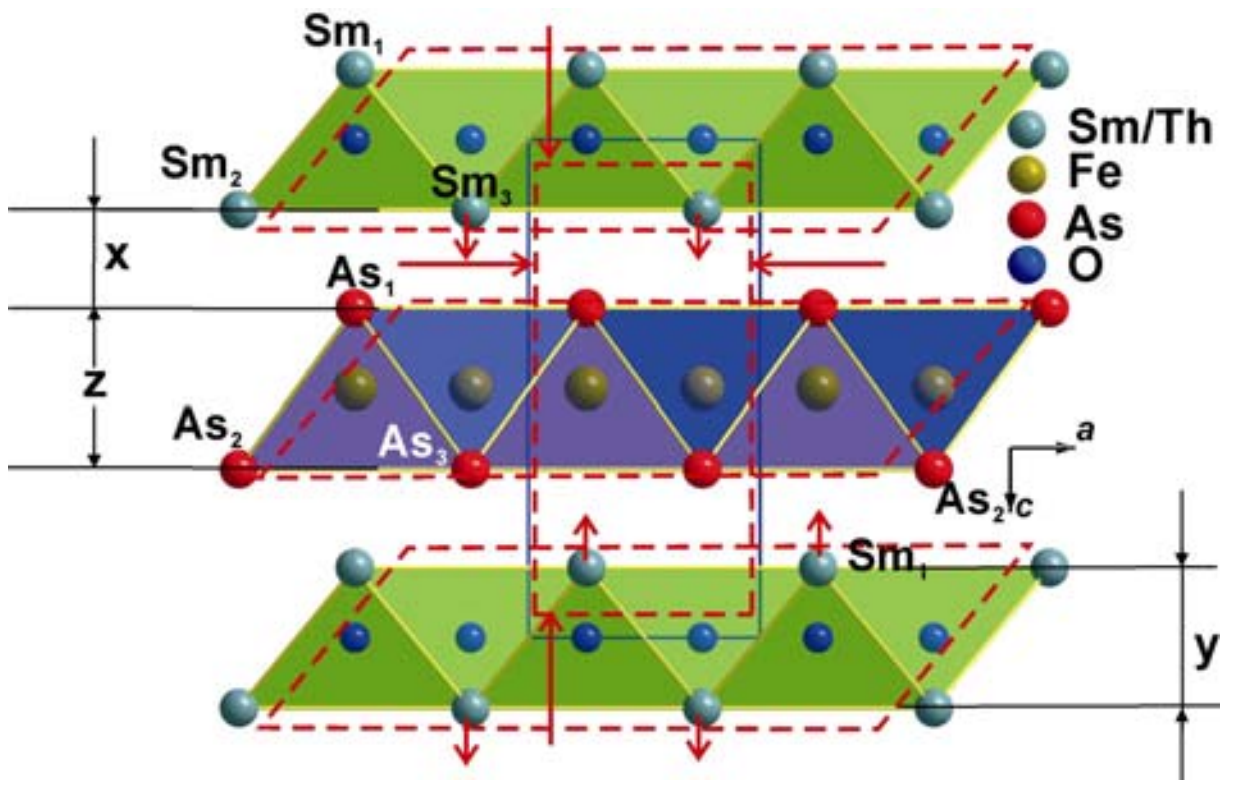

**(a)**

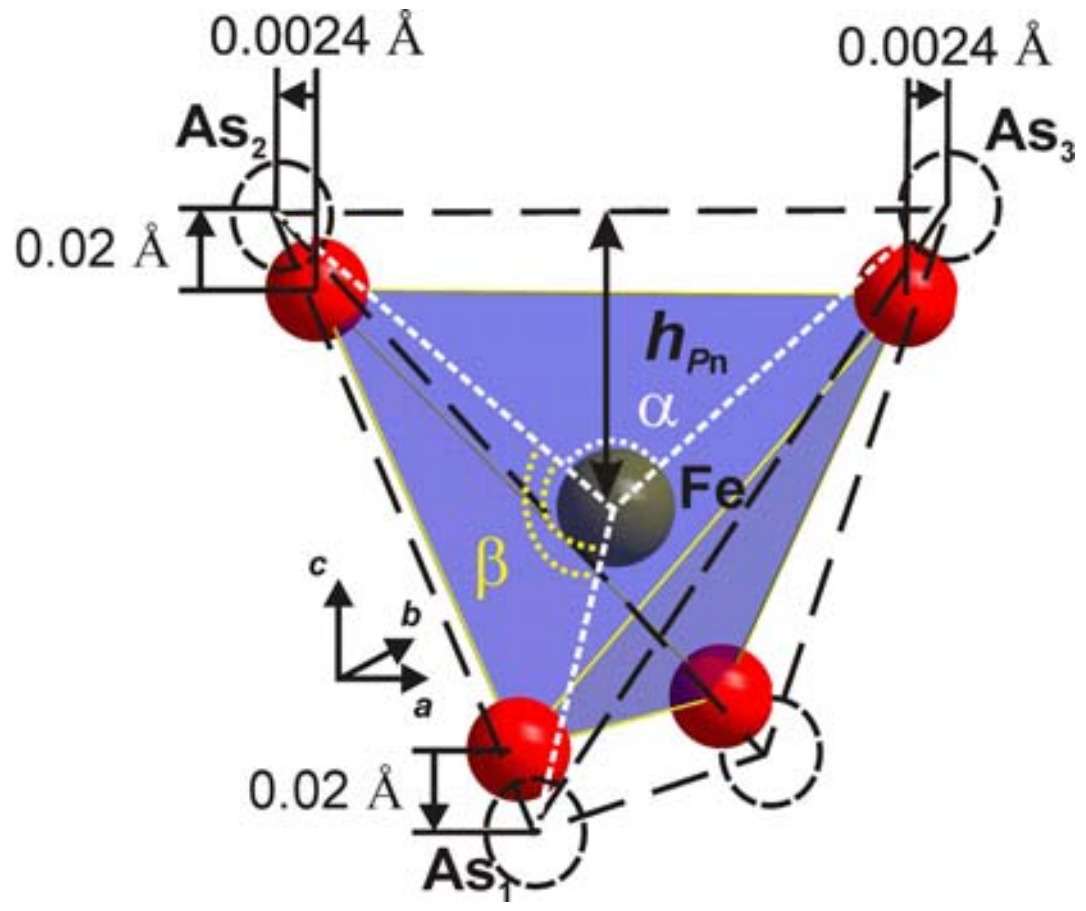

**(b)**

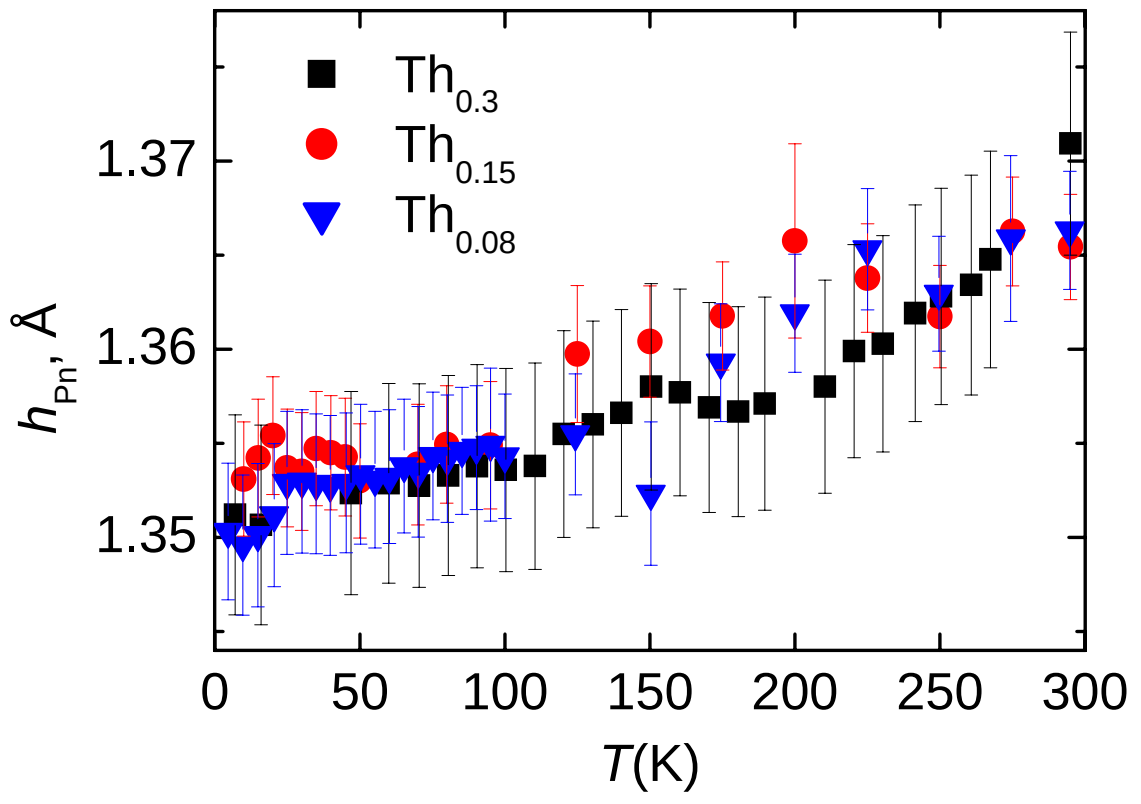

**(c)**

**FIG. 5.** (Color online), (a) Schematic representation of the $Sm_{1-x}Th_xFeAsO$ lattice fragment and its changes (red dotted lines) with substitution of Sm by Th. (b) Schematic representation of the thermal expansion of $FeAs_4$ tetrahedron for $Sm_{0.7}Th_{0.3}FeAsO$ at 15 K (solid lines and circles) and 295 K (dotted lines and circles). (b) Pnictogen height ($h_{Pn}$) as a function of temperature for the three $Sm_{1-x}Th_xFeAsO$ compositions.

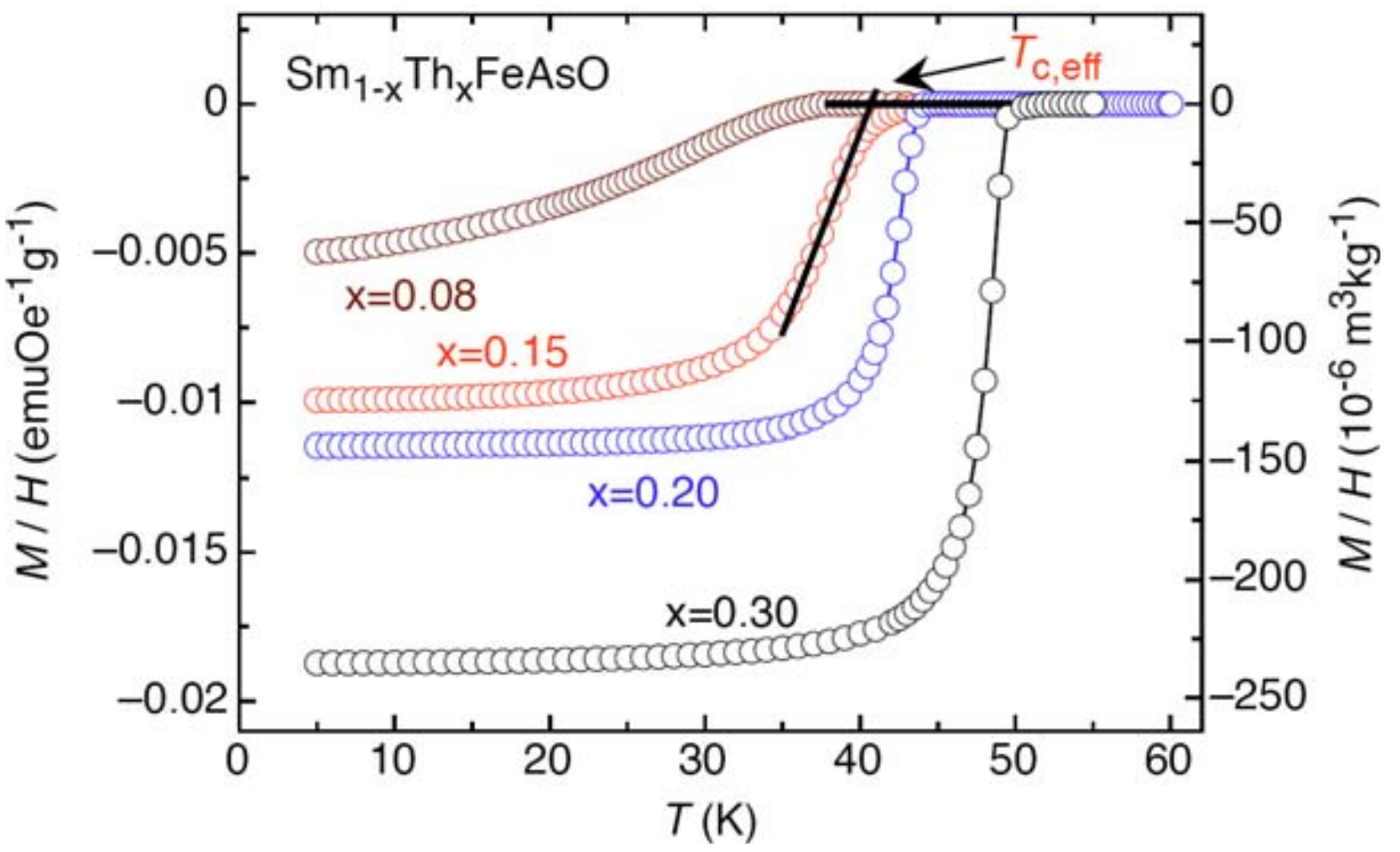

**FIG. 6.** (Color online) Temperature dependence of the magnetic susceptibility for $Sm_{1-x}Th_xFeAsO$ samples measured at 0.5 mT, after cooling in zero field (ZFC). The determination of $T_{c,eff}$ is illustrated.

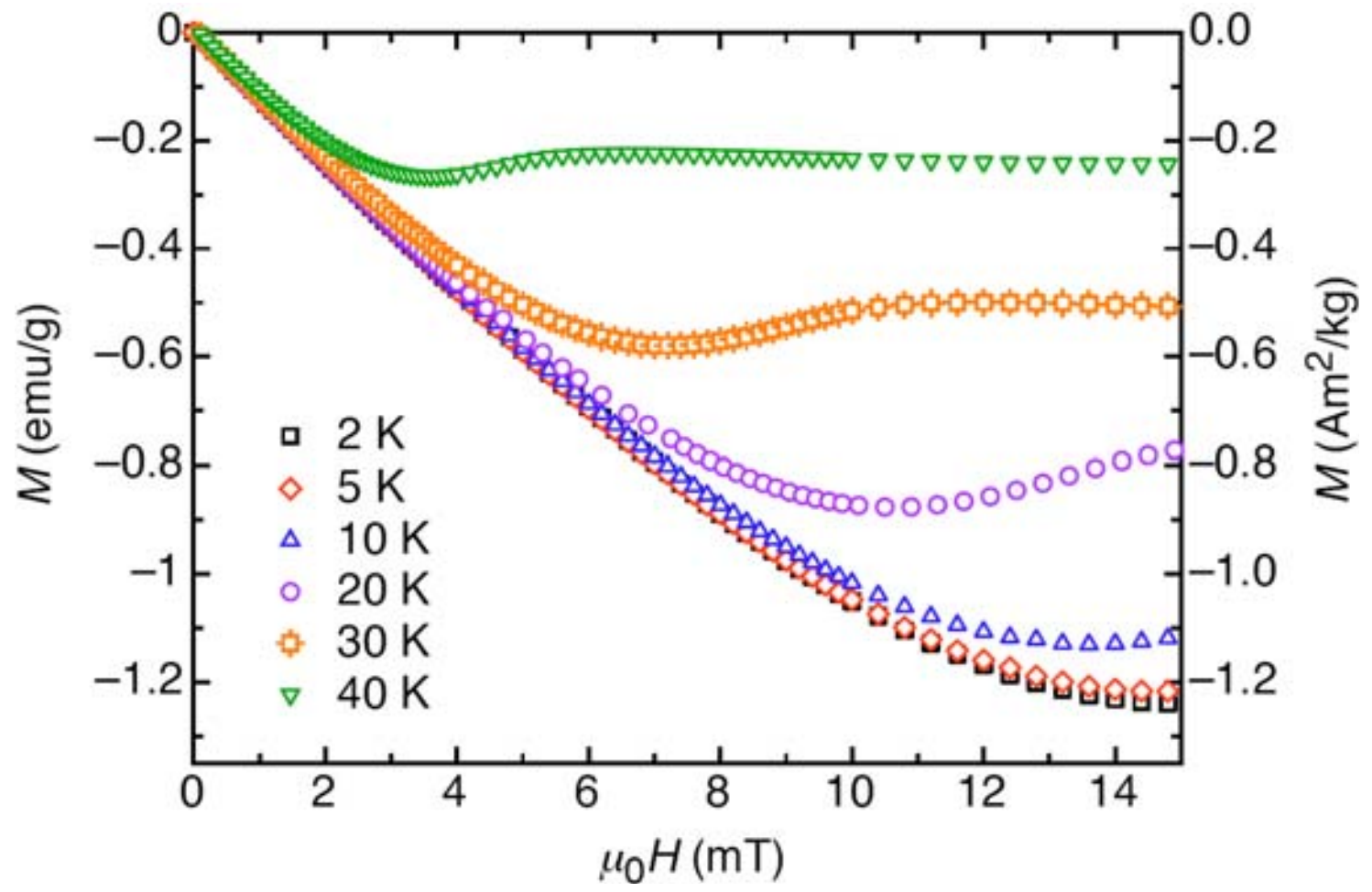

**FIG. 7.** (Color online) Field dependence of the initial magnetization for polycrystalline $Sm_{0.7}Th_{0.3}FeAsO$.

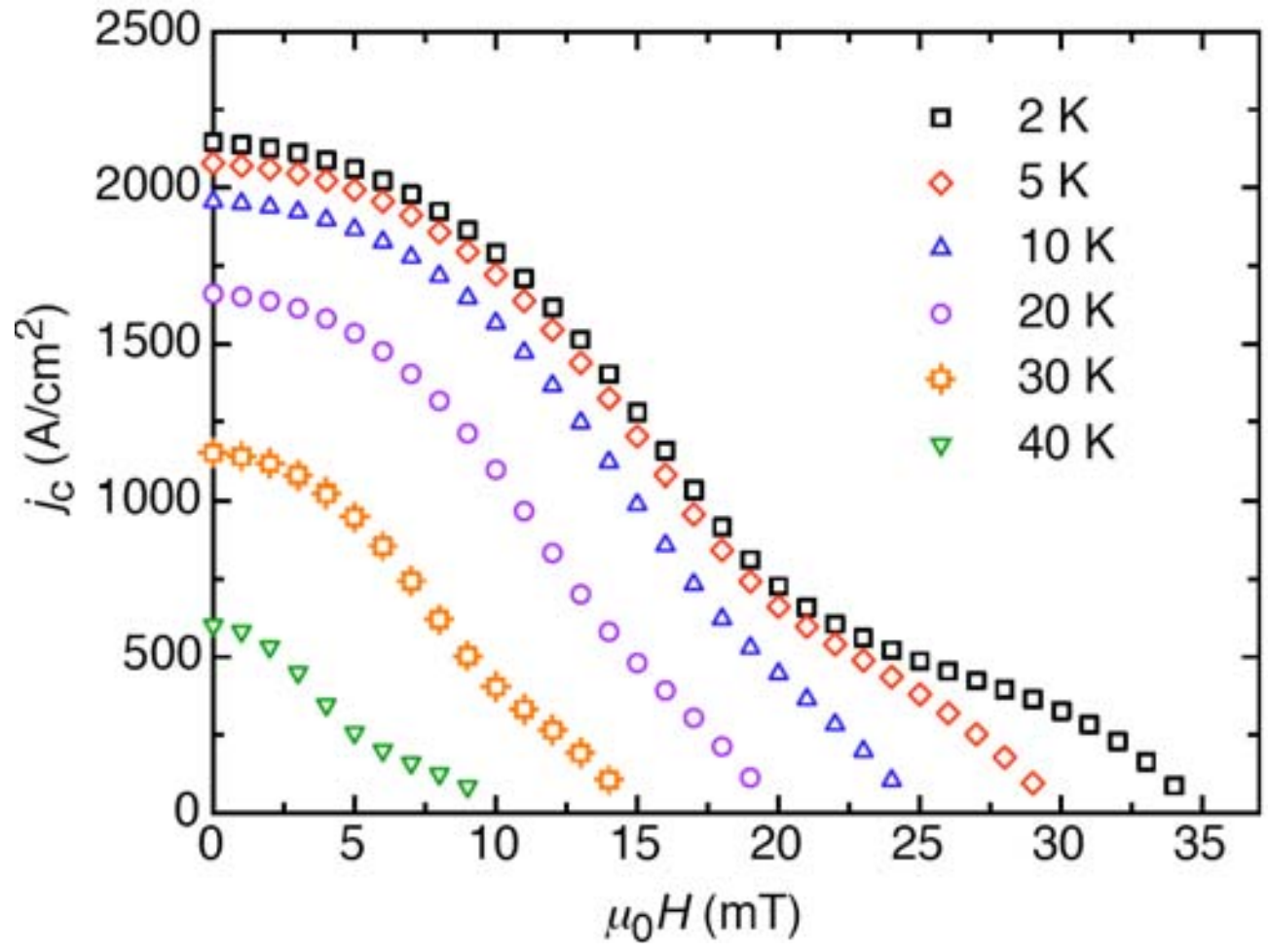

**FIG. 8.** (Color online) Field dependence of the intergrain critical current at various temperatures for polycrystalline $Sm_{0.7}Th_{0.3}FeAsO$.

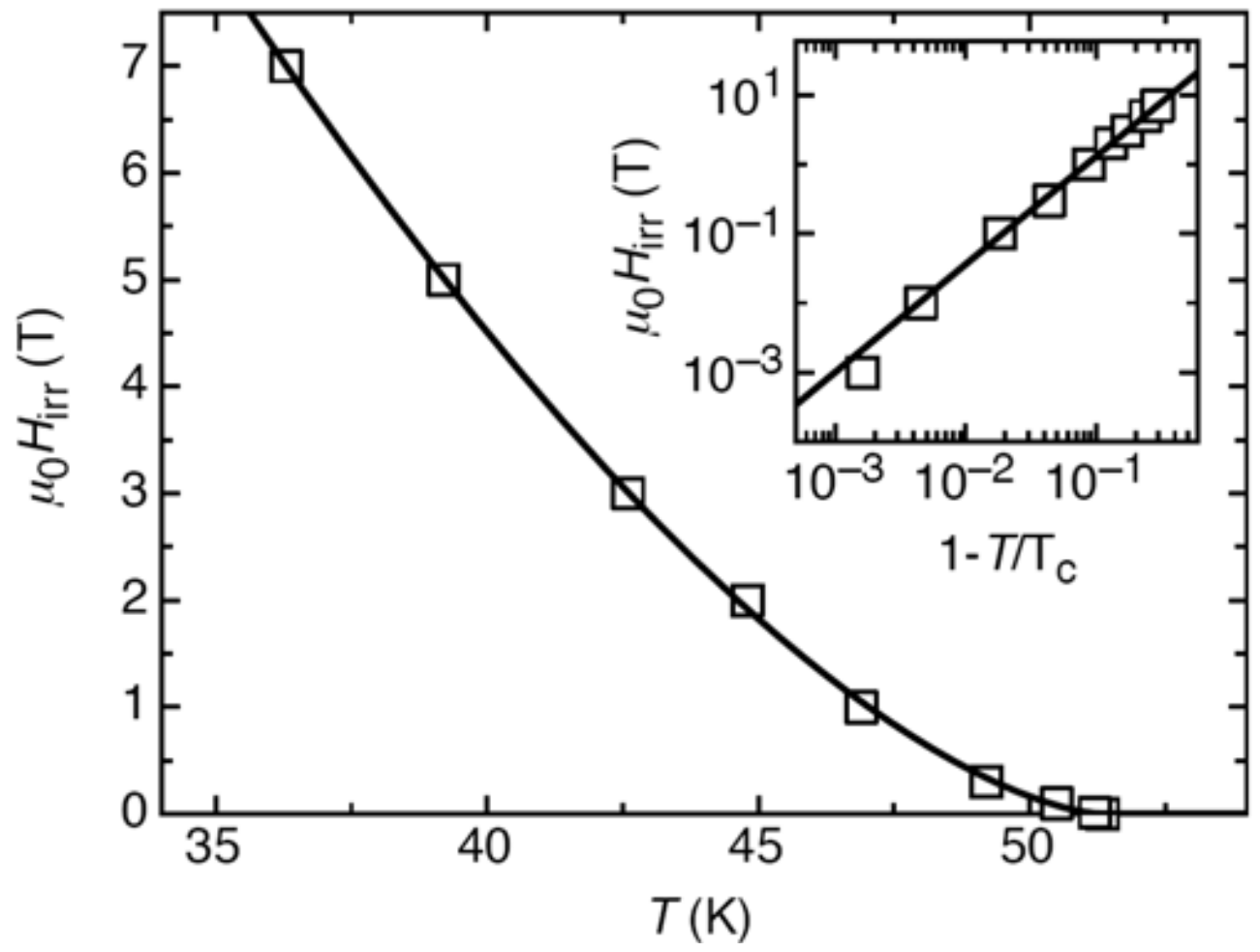

**FIG. 9.** Temperature dependence of the irreversibility field $H_{irr}$ for polycrystalline $Sm_{0.7}Th_{0.3}FeAsO$. The irreversibility line is approximated well by a power-law temperature dependence with an exponent $n \approx 3/2$ (solid line), as discussed in the text. The inset presents the log-log $H_{irr}$ vs. $(1-T/T_c)$ dependence.

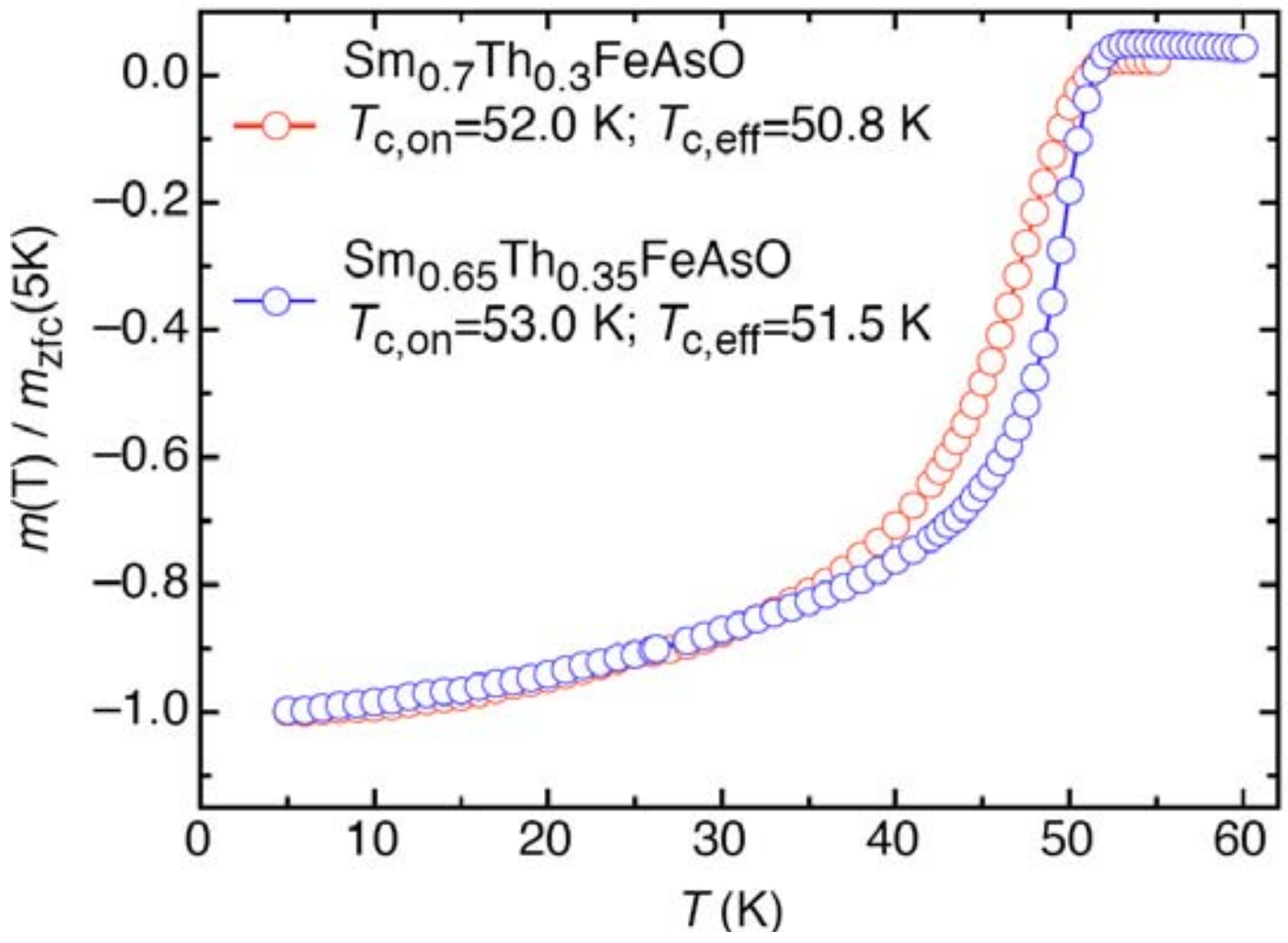

**FIG 10.** (Color online) The normalized diamagnetic signal measured on small lump pieces of $Sm_{1-x}Th_xFeAsO$ ($x$ = 0.3 and 0.35) extracted from the products of single crystal growth.

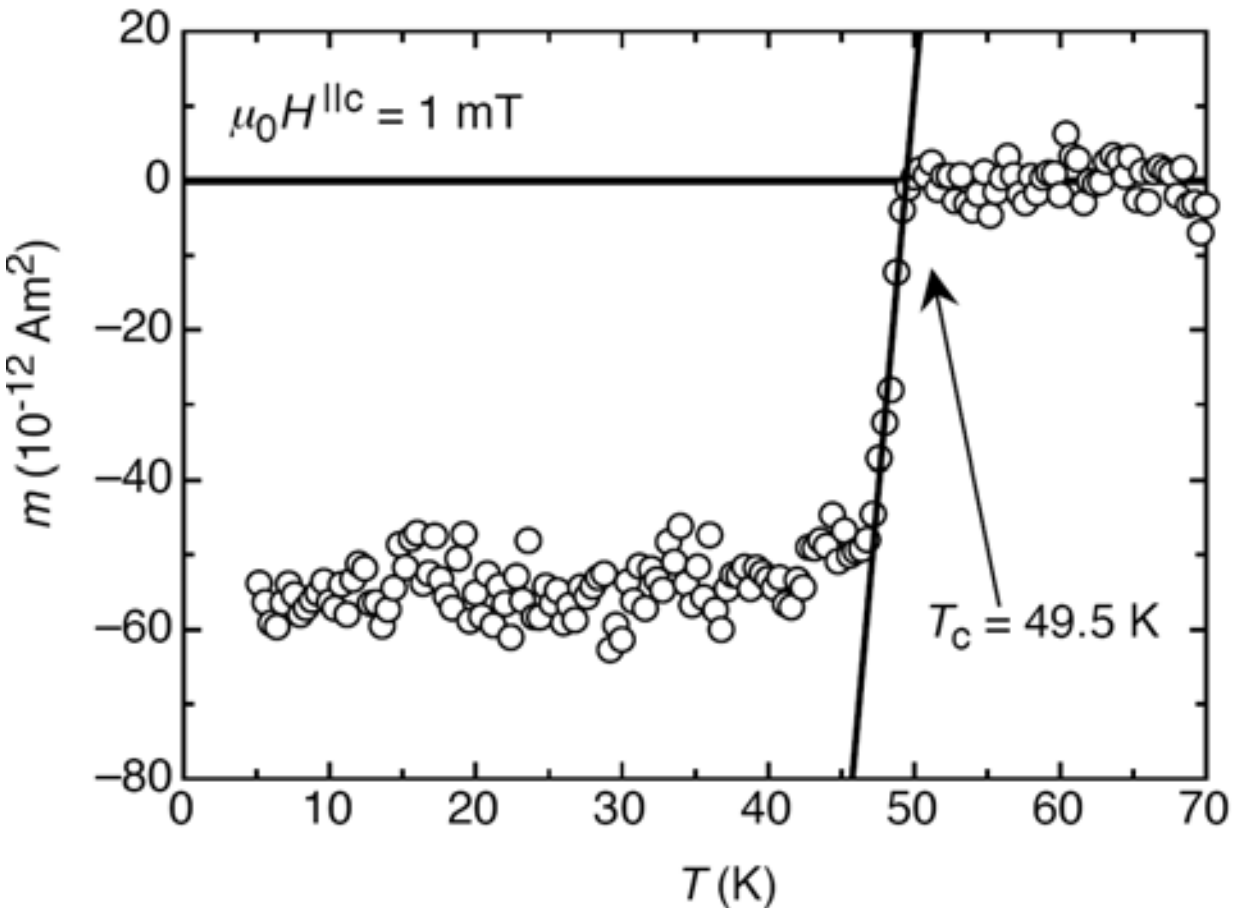

**FIG. 11.** Temperature dependence of the magnetic moment measured on one small single crystal of $Sm_{0.89}Th_{0.11}FeAsO$. The moment was measured in the zero field cooling mode with an applied field of 1 mT parallel to the *c*-axis.

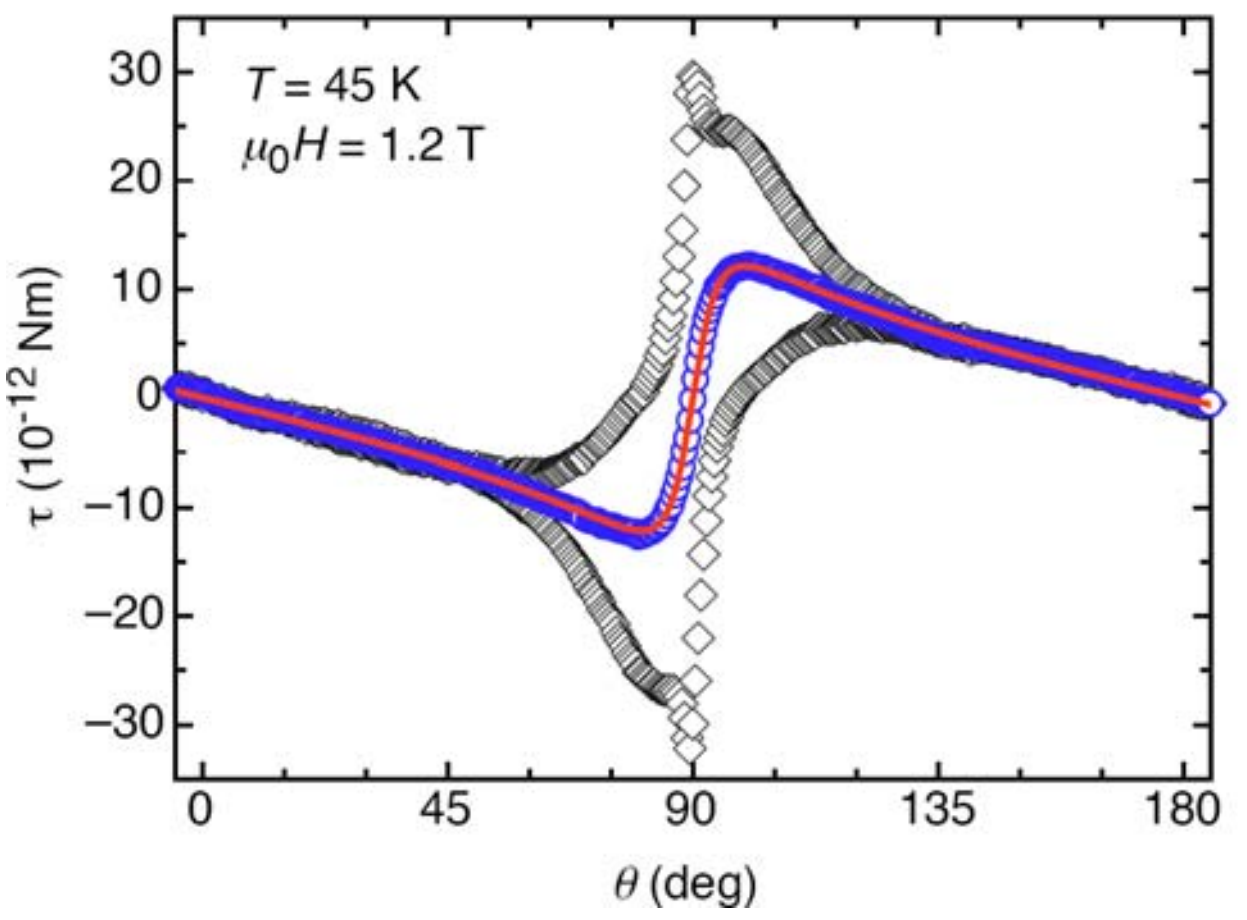

**FIG. 12.** (Color online) Angular dependence of the raw (black open diamonds) and the reversible (open circles) torque data of a $Sm_{0.89}Th_{0.11}FeAsO$ single crystal at $T$ = 45 K and in a magnetic field of 1.2 T. Close to the *ab*-plane, clear indications of a lock-in effect of vortices are observed. The solid curve represents a fit of $\tau(\theta)$, given in Eq. (2), to the reversible torque data (open circles).

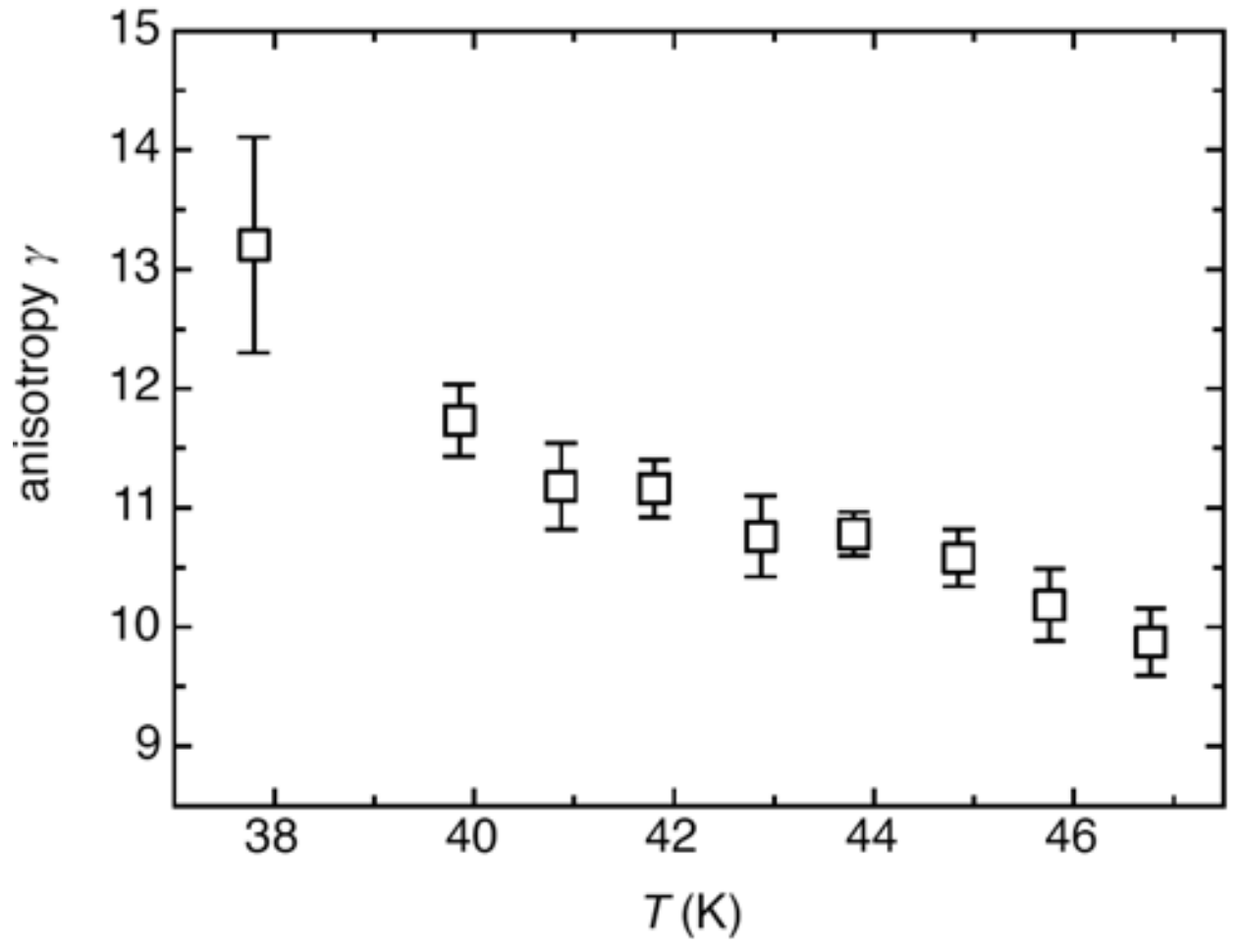

**FIG. 13.** Temperature dependence of the anisotropy parameter $\gamma_\lambda$ obtained from the fits torque data (e.g. Fig. 13).

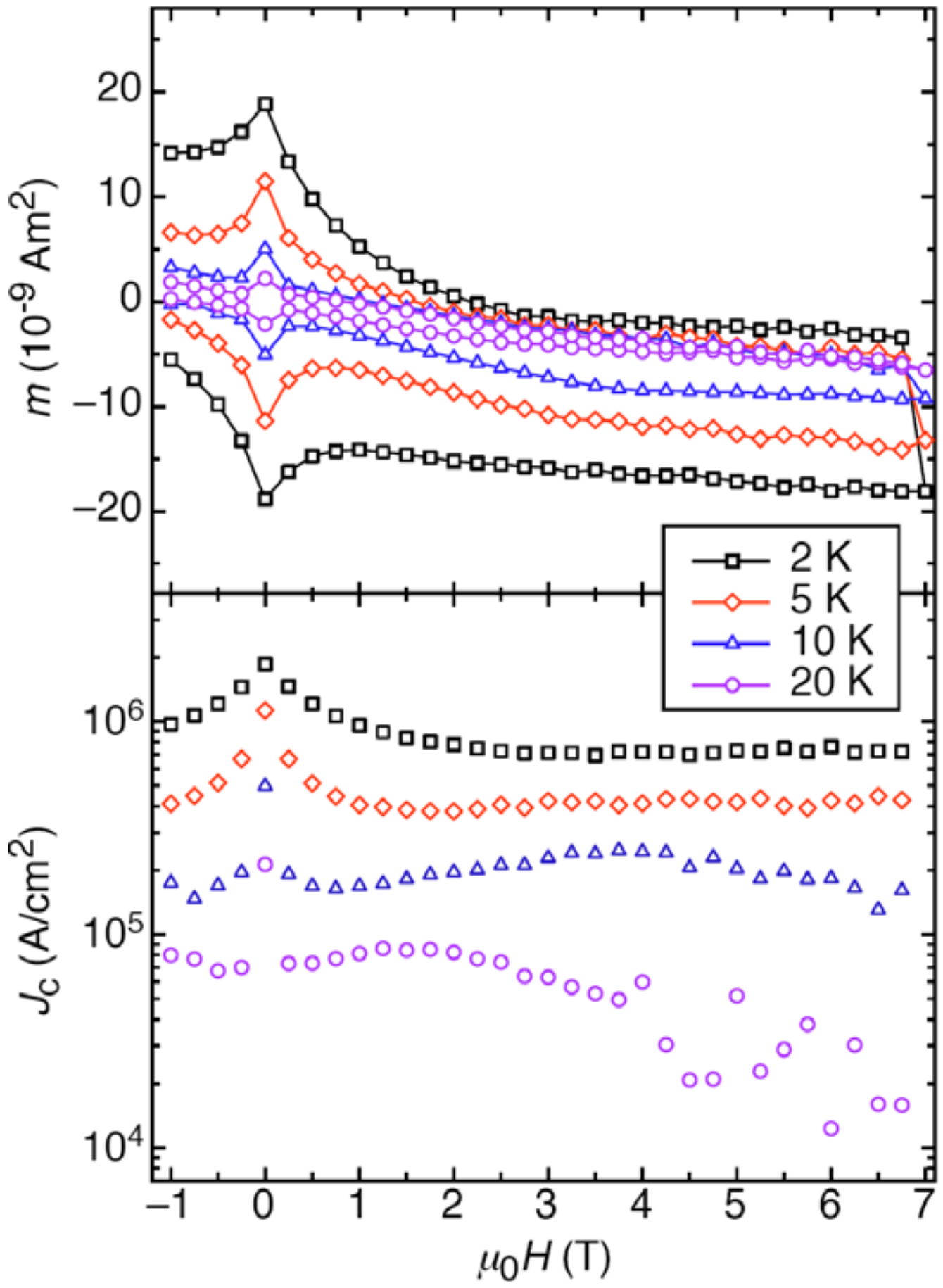

**FIG. 14.** (Color online) Upper panel. Magnetic hysteresis loops measured on a single crystal at 2 K, 5 K, 10 K, and 20 K in a field up to 7 T parallel to the *c*-axis. Lower panel. Critical current density calculated from the width of the hysteresis loops.

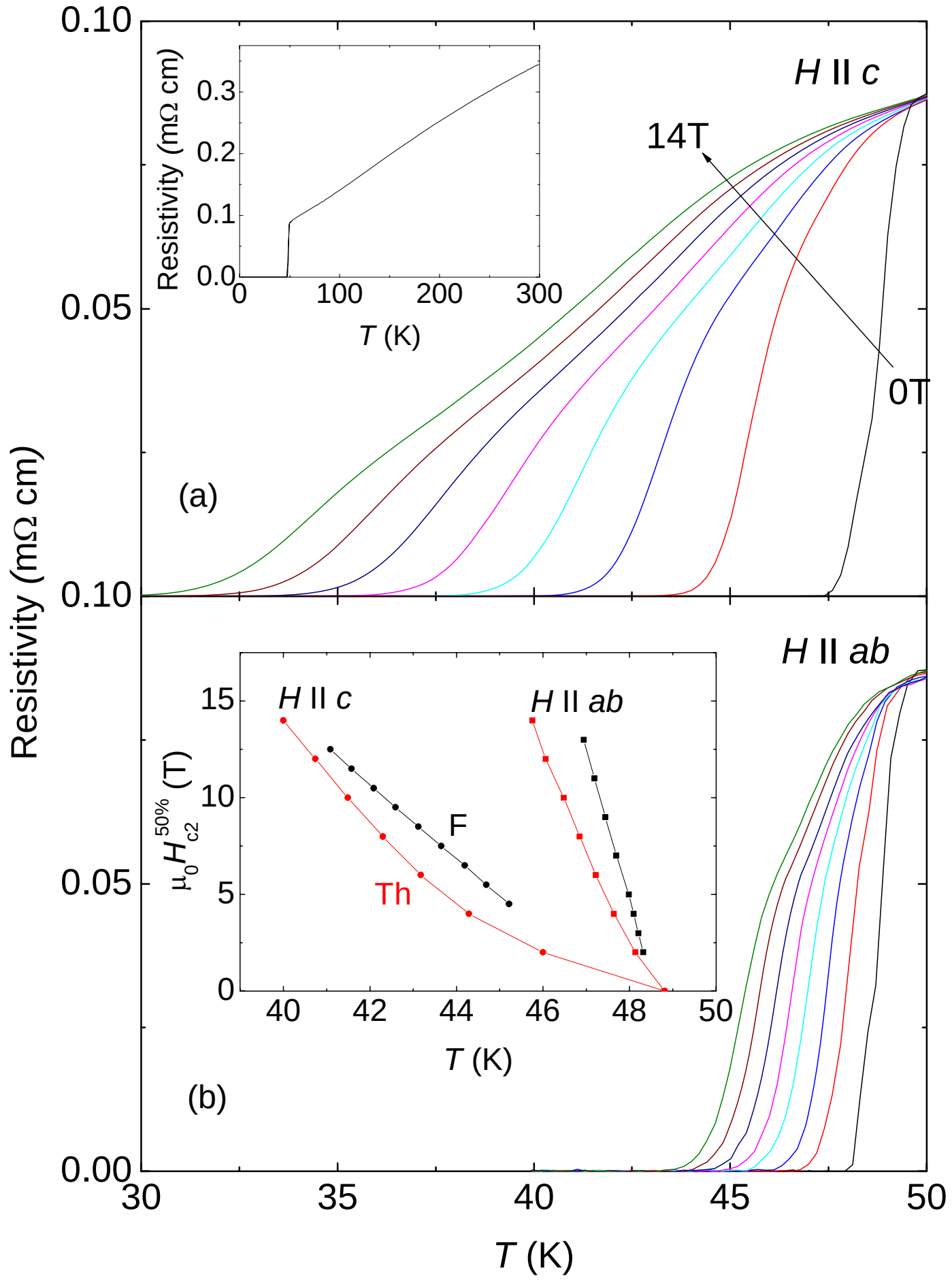

**FIG. 15** (Color online) Temperature dependences of the resistivity for a $Sm_{0.89}Th_{0.11}FeAsO$ single crystal measured with the field applied parallel to the ($Fe_2As_2$) layers ($H||ab$) (a) and perpendicular to them ($H||c$) (b), at the various magnetic fields from 0 to 14 T (0, 2, 4, 6, 8, 10, 12, 14 T). Inset of a) shows the resistivity in the temperature range of 2 – 300 K. Inset of b) shows temperature dependence of the upper critical field with $H||ab$ and $H||c$ for the $Sm_{0.89}Th_{0.11}FeAsO$. To determine $H_{c2}$ the 50% $\rho_n$ criterion was used. The data for $SmFeAsO_{0.7}F_{0.25}$ (nominal composition) crystal were taken from Ref. 38.